\documentclass[%
superscriptaddress,
longbibliography,
 amsmath,amssymb,
 aps, twocolumn,
 longbibliography,
]{revtex4-1}

\usepackage{graphicx}
\usepackage{dcolumn}
\usepackage{xr}
\usepackage{braket}
\usepackage{bm}
\usepackage[colorlinks=true,
  allcolors={blue}]{hyperref}

\newcommand{\fgt}{Fe$_3$GeTe$_2$}

\usepackage{changes}

\graphicspath{{Figures/}}
\begin{document}


\title{Electron transport and scattering mechanisms in ferromagnetic monolayer \fgt{}}

\author{Danis I. Badrtdinov}
\affiliation{Institute for Molecules and Materials, Radboud University, Heijendaalseweg 135, NL-6525 AJ Nijmegen, The Netherlands}

\author{Georgy V. Pushkarev}
\affiliation{Theoretical Physics and Applied Mathematics Department, Ural Federal University, 620002 Yekaterinburg, Russia}

\author{Mikhail I. Katsnelson}
\affiliation{Institute for Molecules and Materials, Radboud University, Heijendaalseweg 135, NL-6525 AJ Nijmegen, The Netherlands}

\author{Alexander N. Rudenko}
\email{a.rudenko@science.ru.nl}
\affiliation{Institute for Molecules and Materials, Radboud University, Heijendaalseweg 135, NL-6525 AJ Nijmegen, The Netherlands}

\date{\today}

\begin{abstract}
We study intrinsic charge-carrier scattering mechanisms and determine their contribution to the transport properties of the two-dimensional ferromagnet \fgt{}. We use state-of-the-art first-principles calculations combined with the model approaches to elucidate the role of the electron-phonon and electron-magnon interactions in the electronic transport. Our findings show that the charge carrier scattering in \fgt{} is dominated by the electron-phonon interaction, while the role of magnetic excitations is marginal. At the same time, the magnetic ordering is shown to effect essentially on the electron-phonon coupling and its temperature dependence. This leads to a sublinear temperature dependence of the electrical resistivity near the Curie temperature, which is in line with experimental observations. The room temperature resistivity is estimated to be $\sim$35 $\mu \Omega \cdot$cm which may be considered as an intrinsic limit for monolayer \fgt{}. 

\end{abstract}

\maketitle


\section{\label{sec1}Introduction}
The interest to two-dimensional (2D) systems is driven by the continuous progress in the development of novel technologies involving miniaturization of electronic devices and low-energy consumption. A special place in this search is devoted to 2D magnetic materials, which became  the subject of active studies after the exfoliation of van der Waaals magnets such as CrI$_3$~\cite{CrI3_Huang2017}, Cr$_2$Ge$_2$Te$_6$~\cite{Cr2Ge2Te6_Gong2017}, Fe$_3$GeTe$_2$ (FGT)~\cite{FGT_Deiseroth2006, FGT_Deng2018, FGT_Fei2018, FGT_Li2018}, etc. Magnetic properties of these systems demonstrate high tunability (e.g. by external electrical field~\cite{Huang2018} and environmental screening~\cite{Soriano2021}) promising for technological application, as well as serve as an excellent platform for probing the magnetic interactions in low dimensions. 

Most of the known 2D magnets are insulating or semiconducting, which limits our understanding of their transport properties and the underlying physical mechanisms. In contrast, FGT has a number of unique characteristics  
since it combines a metallic behavior, needed for the realization of controllable transport~\cite{FGT_Kim2018, FGT_Transport_Roemer2020}, with the ferromagnetic ground  state and comparably high Curie temperature $T_C \simeq $  220 K~\cite{FGT_Deiseroth2006, FGT_Chen2013, FGT_Deng2018}, surviving down to the monolayer limit~\cite{FGT_Fei2018}. One of the most remarkable properties of FGT is the absence of inversion symmetry, allowing for the formation of topologically nontrivial magnetic textures \cite{FGT_DMI_Laref2020, FGT_Ado2021}. A hexagonal lattice of skyrmions is recently observed in this system~\cite{FGT_Ding2020, FGT_skyrmions_Meijer2020, FGT_skyrmions_Park2021}, which might be important for spintronics applications. While a considerable attention is paid to skyrmions,
the electron transport and scattering mechanisms in FGT remain unclear from the microscopic point of view. 

In comparison to nonmagnetic 2D systems, the charge carrier scattering in conductive magnets is not limited to the impurities or phonons, but may include scattering by spin fluctuations \cite{Friedrich2019, Raquet2002}, providing essential contribution to the transport characteristics and/or give rise to the qualitatively new effects (e.g., Kondo effect). Moreover, temperature dependence of the spin-polarized electronic structure in magnetically ordered systems might play a role for the conventional sources of scattering (e.g., phonons).  The first-principles theory of electron-phonon interactions is well established~\cite{Giustino2017}, and has been routinely applied to study transport properties of nonmagnetic 3D and 2D materials~\cite{Ponce2018, Sohier2018, Lugovskoi2019, Ponce2020, Rudenko2020}. Strictly speaking, a theoretical description of the electron-phonon scattering in 2D is more involved due to the presence of flexural phonon modes \cite{Graphene_book}, allowing for multiple phonon scattering \cite{Graphene_book,Rudenko2016,PhysRevB.93.155413}, yet the single-phonon formulation turns out to be sufficient even for a quantitative description of the charge carrier transport in most of the cases \cite{Rudenko2019}. Generalization of the first-principles scattering theory to magnetic materials is not straightforward for at least two reasons: (i) Unequal and temperature-dependent contribution of majority and minority electronic states; (ii) The presence of additional scattering channels such as collective spin excitations (magnons) and spin inhomogeneities. Despite a notable progress in this direction being made in recent years, the proposed theories for the first-principles description of electron-magnon interactions \cite{Friedrich2019, Nabok2021} are limited to zero temperature, i.e. not readily applicable to study transport characteristics. Alternatively, there are well-established theories based on the model description of magnetism~\cite{Irkhin1989,Katsnelson2008,Irkhin2007_book}, which are sufficient to capture the corresponding effects qualitatively and, in many cases, quantitatively.

In this work, we perform a systematic study of the charge carrier transport in ferromagnetic monolayer \fgt{} using first-principles calculations combined with magnetic models. We analyze electron-phonon scattering and interpolate its magnitude between the ferromagnetic and paramagnetic states. In addition, we explicitly consider electron-magnon interaction and estimate its contribution to the temperature-dependent scattering rate. Compared to the spin-resolved electron-phonon coupling constant $\lambda \sim 0.2-0.5$, the electron-magnon coupling constant is found to be smaller by an order of magnitude. The main contribution of the electron-magnon scattering to the resistivity is observed around the Curie temperature, resulting in a marginal resistivity enhancement. On the other hand, interpolation of the electron-phonon scattering between the ferromagnetic and paramagnetic phases leads to a pronounced modification of the resistivity dependence on temperature. Close to the Curie temperature, we obtain a sublinear temperature dependence, which in line with the experimental observations, reporting deviations from the linear dependence in this regime \cite{FGT_Transport_Feng2022, FGT_transport_Chen2013, FGT_transport_Liu2018, Deng2018, FGT_Kim2018}. The resulting lower limit for the room-temperature resistivity in monolayer \fgt{} is found to be around 35 $\mu\Omega\cdot$cm.

The rest of the paper is organized as follows. Sec.~\ref{sec:Method} covers the methods and approximations used in this work. In  Sec.~\ref{sec:Results}, we  present the results and discussion on the spin-polarized electron-phonon couplings in monolayer \fgt{}, phonon- and magnon-mediated scattering rates, as well as the temperature dependence of the electrical resistivity. Finally, Sec.~\ref{sec:Conc} concludes the paper.

\begin{figure}
\centering
\includegraphics[width=0.95\linewidth]{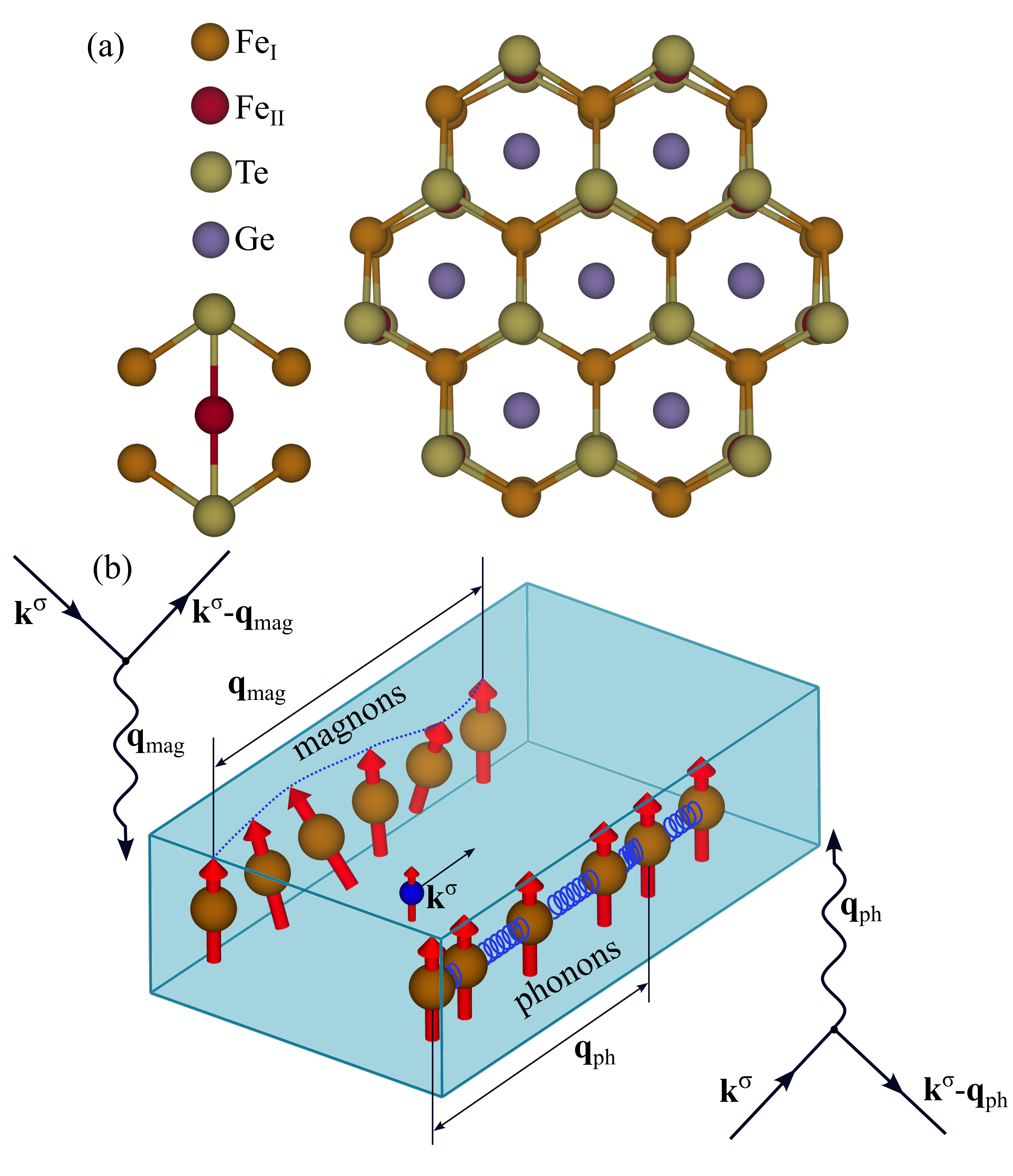}
\caption{(a) Crystal structure of \fgt{}: Side view (left) and top view (right). Fe$_{\rm I}$ and Fe$_{\rm II}$ denote two inequivalent iron atoms. (b) A sketch showing scattering of an electron on phonons and magnons. }
\label{fig:Structure}
\end{figure}

\section{\label{sec:Method}Method}
\subsection{\label{subsec1_1}Technical details}
Electronic structure calculations and structural optimization were performed within plane-wave  based {\sc Quantum Espresso ({\sc qe})}~\cite{QE1,QE2} package utilizing Perdew-Burke-Ernzerhof (PBE) exchange-correlation functional~\cite{PBE} and 2.0.1  version scalar-relativistic norm-conserving pseudopotentials, generated using "atomic" code by A. Dal Corso  v.5.0.99 svn rev. 10869. In these calculations, we use 80 Ry as the plane-wave energy cutoff, a ($16\times16\times1$) $\mathbf{k}$-mesh for the Brillouin zone integration, and $10^{-8}$ eV for the energy convergence criteria. The experimental crystal structure of bulk \fgt{} was used~\cite{Deiseroth2006}, where a vacuum space 20 \AA \, between monolayer replicas in the vertical $z$ direction was introduced to avoid spurious interactions between the periodic supercell images. The effective volume  $\Omega = d \cdot \cal{S}$ with layer thickness $d$ = 8.165~\AA \, of 2D unit cell area $\cal{S}$ = 13.95 \AA$^2$ was used in the calculations. Positions of atoms were allowed to relax  until all the residual force components of each atom were less than  10$^{-4}$ eV/\AA. In all calculations we consider the ferromagnetic (FM) order of spins. From the obtained electronic structure, maximally localized Wannier functions~\cite{marzari1997} were constructed using the {\sc wannier90} package~\cite{pizzi2020} projecting onto the valence Fe($3d$) and Te($5p$) states, which were used in the calculations presented below. 
Phonon spectra were calculated using density functional perturbation theory (DFPT)~\cite{Baroni2001} implemented in the {\sc qe} package. In these calculations, we use a ($4\times4\times1$) $\mathbf{q} -$mesh and a $10^{-14}$ eV as the self-consistency threshold.

\subsection{\label{subsec1_2}Electron transport}
In the semi-classical Boltzmann theory, the in-plane $xx$ component of the conductivity tensor has the form~\cite{Ziman1964_book,Ziman2001_book}:
\begin{equation}
\sigma^\sigma_{xx} = -\frac{e^2}{\Omega}\sum_{n\mathbf{k}} \frac{\partial f^\sigma_{n\mathbf{k}} }{\partial \varepsilon^\sigma_{n\mathbf{k}}}  \, \tau^\sigma_{n\mathbf{k}} [v^x_{n\mathbf{k} \sigma }]^2,
\label{eq:Conductivity}
\end{equation}
with the momentum-dependent scattering rate $\tau_{n\mathbf{k}}$, which can be related to the imaginary part of the electron self-energy:
\begin{equation}
   1/\tau^\sigma_{n\mathbf{k}} = \frac{2}{\hbar} \mathrm{Im} \Sigma^\sigma_{n\mathbf{k}}.
\end{equation} 
In the expression (\ref{eq:Conductivity}), $v^x_{n\mathbf{k}\sigma } = \partial \varepsilon^\sigma_{n\mathbf{k}}/ \partial (\hbar k_x)$ is the $x$ (in-plane) component of the group velocity for band $n$ and wave-vector $\mathbf{k}$, 
 and $f^\sigma_{n\mathbf{k}}$ is the Fermi occupation function for the electron states with energy $\varepsilon^\sigma_{n\mathbf{k}}$, $f^\sigma_{m  \mathbf{k}} = (\exp[(\varepsilon^\sigma_{m  \mathbf{k}}-\varepsilon_F)/k_BT] + 1)^{-1}$, where $\varepsilon_F$ is the Fermi energy.
 We note that the expression (\ref{eq:Conductivity}) is applicable at not too low temperatures where vertex corrections, that is, difference between transport and single-electron relaxation time becomes important \cite{Ziman1964_book,Ziman2001_book,Graphene_book}. 

\textit{Phonon}-mediated scattering was analyzed via calculating  electron-phonon interaction. For this purpose  we use  {\sc EPW}~\cite{Ponce2016} code, which takes the advantage of  Wannier functions based interpolation scheme~\cite{Giustino2017}. Initial version of the code was modified to treat the spin channels independently.  The electron self-energy of this interaction in Migdal approximation has the following form:
\begin{equation}
 \begin{aligned}
 & \Sigma^\sigma_{n \mathbf{k}}(\omega, T)  =  \sum_{m \,\mathbf{q} \nu } |g^\sigma_{mn, \nu}(\mathbf{k}, \mathbf{q})|^2  \times   \\
  & \times   \left[ \frac{b_{\mathbf{q}\nu} + f^\sigma_{m  \mathbf{k + q}}}{\omega - \varepsilon^\sigma_{m  \mathbf{k + q}} + \hbar  \omega_{\mathbf{q \nu}} - i\eta}  + \frac{b_{\mathbf{q}\nu} +1 - f^\sigma_{m  \mathbf{k + q}}}{\omega - \varepsilon^\sigma_{m  \mathbf{k + q}} - \hbar  \omega_{\mathbf{q \nu}} - i\eta} \right]  ,  
\label{eq:Selfen_phonon}
\end{aligned}
\end{equation}
where $b_{\mathbf{q}\nu} = (\exp[\hbar \omega_{\mathbf{q \nu}}/k_BT] - 1)^{-1}$ corresponds to the Bose occupation function for phonons with wave vector $\mathbf{q}$, mode index $\nu$ and frequency $\omega_{\mathbf{q \nu}}$.  The electron-phonon matrix elements contain the information about the derivative of self-consisted spin dependent electronic potential $\partial_{\mathbf{q}\nu} V^\sigma$ in the basis of Bloch functions $\psi^\sigma_{n \mathbf{k}}$:   
\begin{equation}
  g^\sigma_{mn, \nu}(\mathbf{k}, \mathbf{q}) =  \sqrt{  \frac{\hbar}{2m_{0}\omega_{\mathbf{q \nu}}}} \braket{\psi^\sigma_{m \mathbf{k+q}} \vert \partial_{\mathbf{q}\nu}V^\sigma \vert \psi^\sigma_{n \mathbf{k}}}. 
\end{equation}

The electron-phonon coupling constant for each phonon mode $\nu$ with wave vector $\mathbf{q}$ is given by
\begin{equation}
\lambda^\sigma_{\mathbf{q} \nu} = \frac{1}{N_F^\sigma \omega_{\mathbf{q} \nu}} \sum_{mn \mathbf{k}} \vert g^\sigma_{mn, \nu}(\mathbf{k}, \mathbf{q}) \vert ^2 \delta({\varepsilon^\sigma_{n\mathbf{k}}}) \delta({\varepsilon^\sigma_{m\mathbf{k+q}}}), 
\label{eq:Lambda}
\end{equation}
where $N_F^\sigma$ is the electron density of states (DOS) for spin $\sigma$ at the Fermi level. 

\textit{Magnon}-mediated scattering is estimated from the electron-magnon interaction. The corresponding self-energy in case of a ferromagnetic order can be calculated in the spirit of the $s-d$ model~\cite{Irkhin1989,Katsnelson2008,Irkhin2007_book} as: 
\begin{equation}
 \begin{aligned}
& \Sigma^\uparrow_{n \mathbf{k}}(\omega, T)   =  2I^2\braket{S_z}  \sum_{\mathbf{q} \nu }  \frac{b_{\mathbf{q}\nu} + f^\downarrow_{n  \mathbf{k + q}}}{ \omega - \varepsilon^\downarrow_{n  \mathbf{k + q}}  +  \hbar \omega_{\mathbf{q \nu}} - i\eta  } \\
& \Sigma^\downarrow_{n \mathbf{k}}(\omega, T) =  2I^2\braket{S_z}  \sum_{\mathbf{q} \nu } \frac{b_{\mathbf{q}\nu} + 1 - f^\uparrow_{n  \mathbf{k - q}}}{ \omega - \varepsilon^\uparrow_{n  \mathbf{k - q}}  - \hbar \omega_{\mathbf{q \nu}} - i\eta  }. \\
\label{eq:Selfen_magnon}
\end{aligned}
\end{equation}

Here, $I$ is the electron-magnon interaction constant ($s-d$ exchange parameter) averaged over the Fermi surface
\begin{equation}
I = \frac{1}{2SN_F} \sum_{m \,\mathbf{k} \sigma} (\varepsilon^\uparrow_{m\mathbf{k}} - \varepsilon^\downarrow_{m\mathbf{k}})  \frac{\partial f^\sigma_{m\mathbf{k}} }{\partial \varepsilon^\sigma_{m\mathbf{k}}},    
\end{equation}
and $\braket{S_z}$ and  $\omega_{{\bf q} \nu}$ correspond  to the temperature-dependent magnetization and magnon frequencies, respectively. Here, $N_F$ is the total electron density of states at the Fermi energy, $N_F  =  N^{\uparrow}_F + N^{\downarrow}_F$. In both approximations [Eqs.~(\ref{eq:Selfen_phonon})  and (\ref{eq:Selfen_magnon})], we consider static limit, i.e. $\omega$ = 0 which is justifiable at not too low temperatures. 

\subsection{Temperature-dependent magnetization\label{sec2e}}
In  order to calculate $\braket{S_z}$ and  $\omega_{{\bf q} \nu}$, we consider the following quantum spin model with $S$ = 1:
\begin{equation}
\hat {\mathcal{H}} = \sum_{i>j} J_{ij} \hat {{\bf S}}_{i} \hat {{\bf S}}_{j} - A\sum_{i}{S^2_{z \,i}}.
\label{eq:spin_ham}
\end{equation}
The magnetic exchange interactions $J_{ij}$ are calculated within the local force theorem approach~\cite{liechtenstein1987, Kashin2020} (see Appendix \ref{app:ISO} for details), with the results presented in Table~\ref{tab:Exchange_couplings}. We note that we are not aiming at a precise determination of the Hamiltonian parameters within this study which would require a complicated discussion on possible quantum corrections, etc. Our main purpose is to qualitatively reproduce the temperature-dependent magnetization as well as the magnon spectrum. 

On-site anisotropy parameter $A$ = 0.35 meV/Fe  is calculated as a total energy difference for magnetic moments oriented along the in-plane and out-of-plane directions taking spin-orbit coupling effects into account. 

Equation~(\ref{eq:spin_ham}) allows us to introduce the  spin-wave Hamiltonian, whose eigenvalues correspond to the magnon frequencies $\omega_{{\bf q} \nu}$ \cite{Rusz2005}:
\begin{equation}
\hat{\mathcal{H}}^{SW}_{\mu \nu}(\mathbf{q})  = \left[ \delta_{\mu \nu}[A + \sum_{\chi} J_{\mu \chi}(\mathbf{0})] - J_{\mu \nu}(\mathbf{q}) \right] \braket{S_z},
\label{eq:SW}
\end{equation}
where $J_{\mu \nu}({\bf q})$ are the Fourier transform of the exchange interactions, and the indices $\mu$ and $\nu$ run from 1 to 3 over the Fe atoms in the unit cell. 
In turn, the magnetization $\braket{S_z}$ entering Eq.~(\ref{eq:spin_ham}) is calculated within the Tyablikov’s Green's functions formalism ~\cite{Tyablikov_book, Nolting_book}. The corresponding expression for $S=1$ takes the form
\begin{equation}
  \braket{S_z} = S \frac{1 + 2 \sum_{\mathbf{q} \nu}b_{\mathbf{q}\nu}}{ 1 + 3 \sum_{\mathbf{q} \nu}b_{\mathbf{q}\nu} + 3(\sum_{\mathbf{q} \nu}b_{\mathbf{q}\nu})^2}.
\label{eq:Magnetization}
\end{equation}
Here, the Bose factors $b_{\mathbf{q}\nu}$ depend on the magnon frequencies $\omega_{{\bf q}\nu}$, which are, in turn, obtained by diagonalizing the ($3\times3$) Hamiltonian matrix [Eq.~(\ref{eq:SW})]. In order to obtain $\braket{S_z}$ and $\omega_{{\bf q}\nu}$ simultaneously, we solve Eqs.~(\ref{eq:SW}) and (\ref{eq:Magnetization}) self-consistently. The resulting magnetization $\braket{S_z}$ drops down to zero at the Curie temperature $T_C$.  Above this temperature the system becomes paramagnetic.


\section{\label{sec:Results}Results}
\subsection{Electronic structure}
\begin{figure}
\centering
\includegraphics[width=1\linewidth]{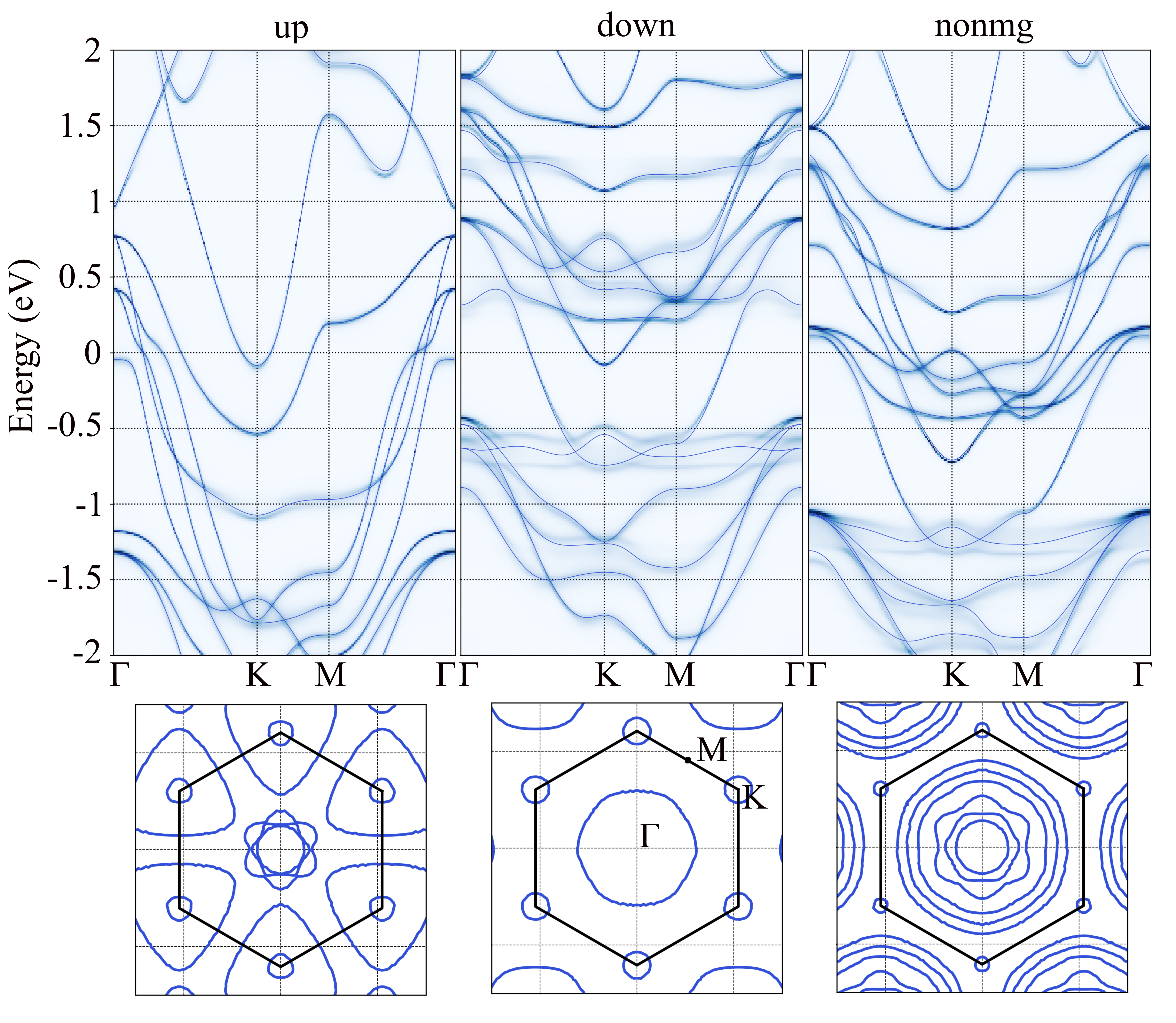}
\caption{(Top) Electron spectral functions $A^{\sigma}_{\mathbf{k}} (\omega, T) = -1/\pi \, \mathrm{Im} [G^\sigma_{\mathbf{k}}(\omega, T)]$ calculated in the presence of the electron-phonon interactions for $T = 100$ K for the states near the Fermi level in monolayer \fgt{}. Original DFT band structure 
is shown by the blue solid line. Zero energy corresponds to the Fermi energy.
(Bottom) The corresponding Fermi contour maps.}
\label{fig:Bands}
\end{figure}
Figure \ref{fig:Bands} shows the electronic structure and Fermi surfaces calculated for monolayer \fgt{}. We explicitly consider electronic states in the ferromagnetic and nonmagnetic phases. The states in the vicinity of the Fermi energy are predominantly formed by the Fe ($3d$) states hybridized with Te ($5p$). The spin-resolved electronic structure in the FM phase displays multiple energy bands crossing the Fermi energy, resulting in the formation of several isolated pockets in the Fermi surface, being in agreement with the experimental data~\cite{FGT_Kim2018, FGT_transport_Zhang2018}. Near the $\Gamma$ point, one can observe a hexagonal-like shaped pocket of the spin-down states, which is to
contribute to the nesting at wave vectors $\mathbf{q = k - k^\prime}$ away from the zone center. Around the K points, one can see circular pockets allowing for transitions near the zone center. The spin-up (majority) Fermi surface is more complicated and it has more possibilities for transition at different $\mathbf{q}$.
In both cases, the momentum transfer processes are expected to occur predominantly near the $\Gamma$ point with discrete regions of ${\bf q}$ points around the nesting wave vectors.
The electronic structure from the nonmagnetic calculations is considerably different. The Fermi surface is composed of circular pockets around the $\Gamma$ point, allowing the transitions in a broad range of $\mathbf{q}$ vectors. We note that a moderate variation of the Fermi level (i.e. the doping effect) can significantly change the observed picture, which will influence the electron-phonon interaction as we will show below. It is also worth noting that the electronic states in monolayer \fgt{} are not strongly affected by the electron-phonon interactions. The spectral functions shown in the top of Fig.~\ref{fig:Bands} do not show any significant renormalization in the vicinity of the Fermi energy, while the linewidths are nearly constant for all relevant bands and ${\bf k}$-points. The DOS calculated for the FM and nonmagnetic states, as well as in the disordered local moment approximation are presented in Appendix \ref{app:dlm} (Fig.~\ref{fig:DLM}).

\subsection{Phonon dispersion}
\begin{figure}
\centering
\includegraphics[width=1\linewidth]{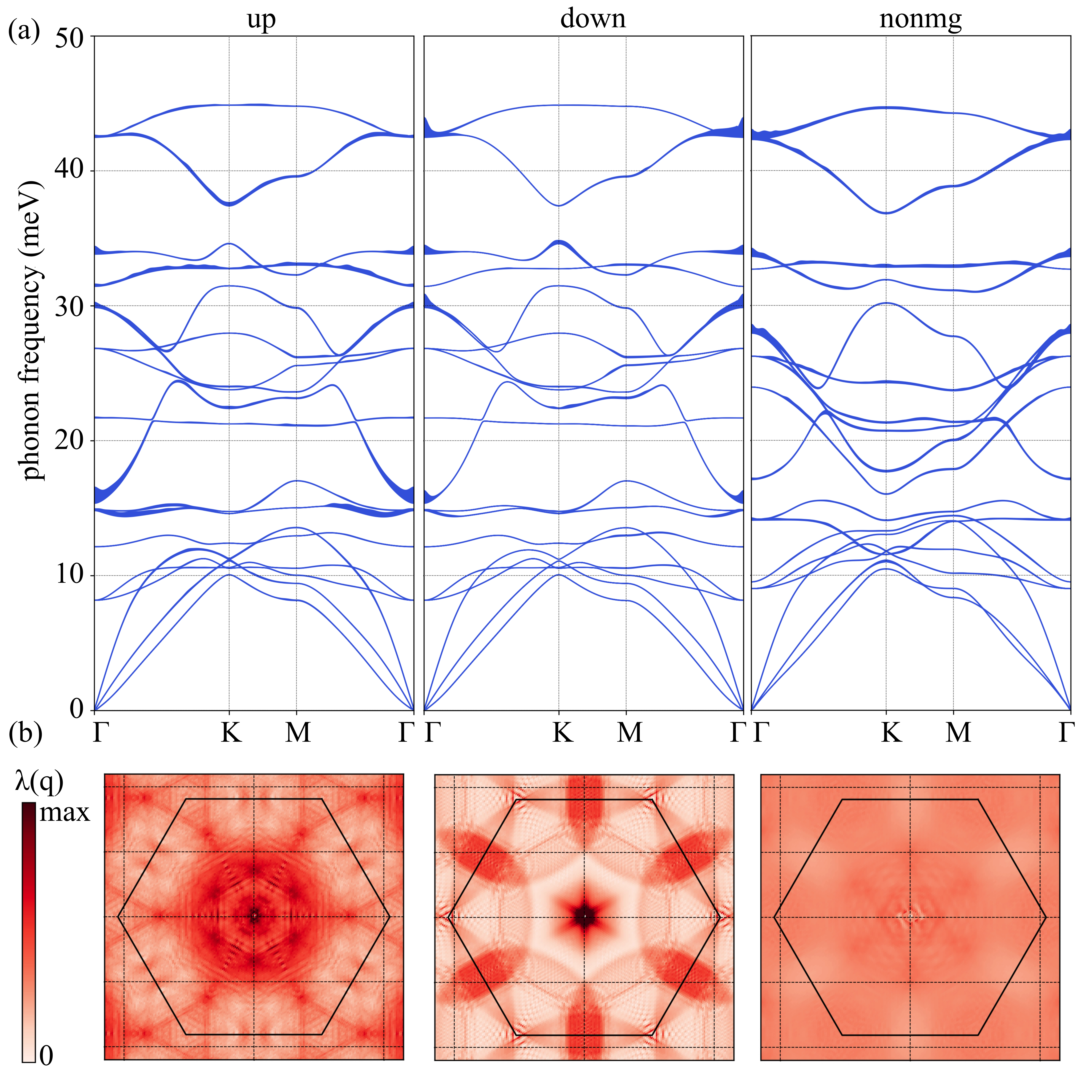}
\caption{(a) Phonon dispersion curves calculated for the ferromagnetic and nonmagnetic phase of monolayer \fgt{}. The linewidth of phonon modes is spin-dependent, proportional to $\mathrm{Im}[\Pi^{\sigma}_{\mathbf{q \nu}}] = \pi N^\sigma_{F} \lambda^{\sigma}_{\mathbf{q \nu}} \omega^2_{\mathbf{q \nu}}$. (b) Momentum-resolved  electron-phonon coupling constant $\lambda^\sigma_{\mathbf{q}} = \sum_\nu \lambda^\sigma_{\mathbf{q \nu}}$.   }
\label{fig:Linewidth}
\end{figure}
The phonon dispersion and the corresponding spin-resolved linewidths are shown in Fig.~\ref{fig:Linewidth} for the ferromagnetic and nonmagnetic calculations. The obtained dispersion of acoustic phonons is typical for 2D systems, demonstrating two linearly-dispersing branches and one flexural out-of-plane mode with a quadratic dispersion around the $\Gamma$ point~\cite{Rudenko2020, Lugovskoi2019}. The optical phonon modes appear at energies from $\sim$8 meV to $\sim$45 meV. Interestingly, one can see a number of nearly flat branches, the most prominent of  which appear near 15, 21, and 32 meV. The phonon modes with energies less than 15 meV correspond predominantly to the vibration of heavy Te atoms, hybridized with Fe and Ge vibration states. The two highest energy modes represent  a hybridized vibration of Fe and Ge atoms, while the rest phonon modes predominantly correspond to the Fe atoms vibration. Overall, the calculated phonon spectra are in good agreement with previously reported data~\cite{Zhuang2016}. The phonon linewidths are somewhat different for the magnetic and nonmagnetic spectra. In particular, for the spin majority states in the ferromagnetic phase, most of the line broadening takes place near the $\Gamma$ point around 15 meV, i.e. within the range of thermal excitations. In the nonmagnentic case, most of the line broadening is observed above 25 meV, thus their excitation is less likely. All this suggests that thermal transport in different magnetic states of FGT must be different, which might be interesting for further studies.

\subsection{Electron-phonon coupling and its interpolation}
We now turn to the electron-phonon interaction in monolayer \fgt{}. As one can see from Fig.~\ref{fig:Linewidth}(b), the ${\bf q}$-resolved electron-phonon interaction constant $\lambda_{\mathbf{q}}$ is essentially different for spin-up and spin-down channels, which is expected from the spin-resolved linewidths shown above. For the spin-down states, the dominant electron-phonon coupling originates from the interaction with long-wavelength phonons, which results in a highly localized maximum of $\lambda_{\mathbf{q}}$ at the  $\Gamma$ point. In the spin-up case, the contributions with $\mathbf{q} > 0$  play a moderate  role, although the maximum is still observed around the $\Gamma$ point. In the nonmagnetic case $\lambda_{\mathbf{q}}$ is distributed practically uniformly, which is in line with the electronic structure and the Fermi contours discussed above.

\begin{figure}
\centering
\includegraphics[width=0.95\linewidth]{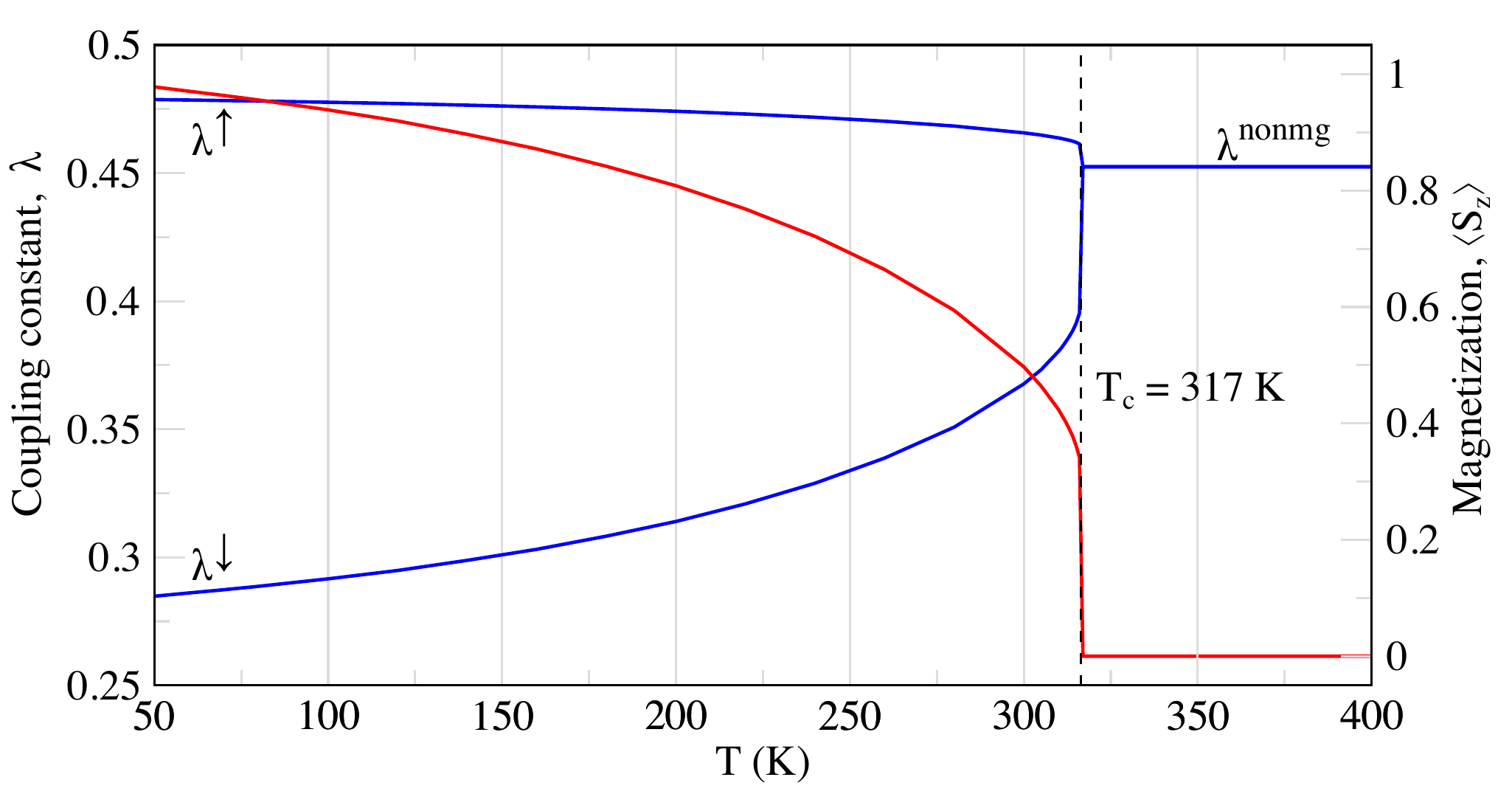}
\caption{Spin-resolved electron-phonon coupling constant  $\lambda^\sigma$ (blue) and magnetization $\braket{S_z}$ (red) calculated for monolayer \fgt{} as a function of temperature using the interpolation scheme discussed in the main text.}
\label{fig:Lambda_T}
\end{figure}

The  integrated  electron-phonon coupling constant $\lambda^\sigma = \sum_{\mathbf{q} \nu}\lambda^\sigma_{\mathbf{q} \nu} $ for spin up and spin down states are found to be 0.50 and 0.26, respectively, i.e. $\lambda^\uparrow/\lambda^\downarrow  \sim  2$. In the nonmagnetic case, we find $\lambda^{\rm nonmg} \simeq$ 0.45. The resulting constant $\lambda^\sigma < 1$, which can be qualified as a moderate electron-phonon interaction. Given that the electron-phonon interaction constant is different for different magnetic states, it is important to take its temperature dependence into account when calculating the transport properties. 

In what follows, we mimic the paramagnetic state of the system by a nonmagnetic solution obtained from the non-spin-polarized DFT calculations described above. Strictly speaking, this assumption is quite strong as it ignores the presence of residual disordered local magnetic moments at $T>T_C$, which may affect the electronic structure and related properties, especially in itinerant magnets. The most appropriate approach to deal with the paramagnetic state in first-principles calculations is to use the method of disordered local moments (DLM) \cite{Gyorffy_1985,Staunton_1985,Pindor_1983,Staunton_1986,PhysRevB.67.235105}. This method requires considering large supercells in the DFT calculations, making DFPT calculations for paramagnetic states prohibitively expensive. However, our DLM calculations of the density of states (DOS) for monolayer \fgt{} (see Appendix \ref{app:dlm} for details) demonstrate that the DLM DOS near the Fermi energy turns out to be comparable to the nonmagnetic DOS. This behavior can be attributed to quenching of the local magnetic moment on the Fe$_{\rm II}$ atoms [see Fig.~\ref{fig:Structure}(a)] in most of the disordered configurations such that $\langle S^2_z \rangle_{{\rm II}} =0$ in the DLM state. Unlike the ground-state FM configuration, the Fe$_{\rm II}$ $d$ states are not expected to be split in the paramagnetic phase, providing a sizeable contribution at the Fermi energy, as in  the nonmagnetic solution.

Based on the argumentation given above, at temperatures $T>T_C$, we assume that the results converge to the nonmagnetic case. We can then approximate the temperature-dependent electron-phonon coupling constant using the interpolation formula: 
\begin{equation}
 \lambda^\sigma (T)  = \lambda_{\rm nonmg} + \left[\lambda^\sigma - \lambda_{\rm nonmg} \right] \frac{\braket{S_z}}{S}.
  \label{eq:lambda_interpolate}
\end{equation}
At $T = 0$ K, the average spin $\braket{S_z}$ equals to the nominal spin $S=1$, yielding $\lambda^\sigma(0) = \lambda^\sigma$. On the other hand, above Curie temperature, we have  $\lambda^\sigma(T > T_C) = \lambda_{\rm nonmg}$. The magnetization is determined self-consistently as described in Sec.~\ref{sec2e}. The resulting magnetization curve is shown in Fig.~\ref{fig:Lambda_T}, from which the Curie temperature $T_C$ = 317 K can be determined. The obtained value overestimates the  experimental temperature $T_C$ $\sim$ 200 K \cite{Deng2018, Fei2018, Verchenko2015, Bin2013}. Nevertheless, obtained magnetization curve $\braket{S_z}$ can be used to interpolate the results between magnetic and nonmagnetic solutions. In Fig.~\ref{fig:Lambda_T}, we plot  the result of the electron-phonon coupling constant interpolation using Eq.~(\ref{eq:lambda_interpolate}). With the increase of temperature the constants $\lambda^\sigma$ for both spin channels smoothly change their values, converging into the nonmagnetic $\lambda_{\rm nonmg}$ = 0.45 result  at $T_C$. 

\subsection{Temperature-dependent scattering rate}

\begin{figure}
\centering
\includegraphics[width=0.95\linewidth]{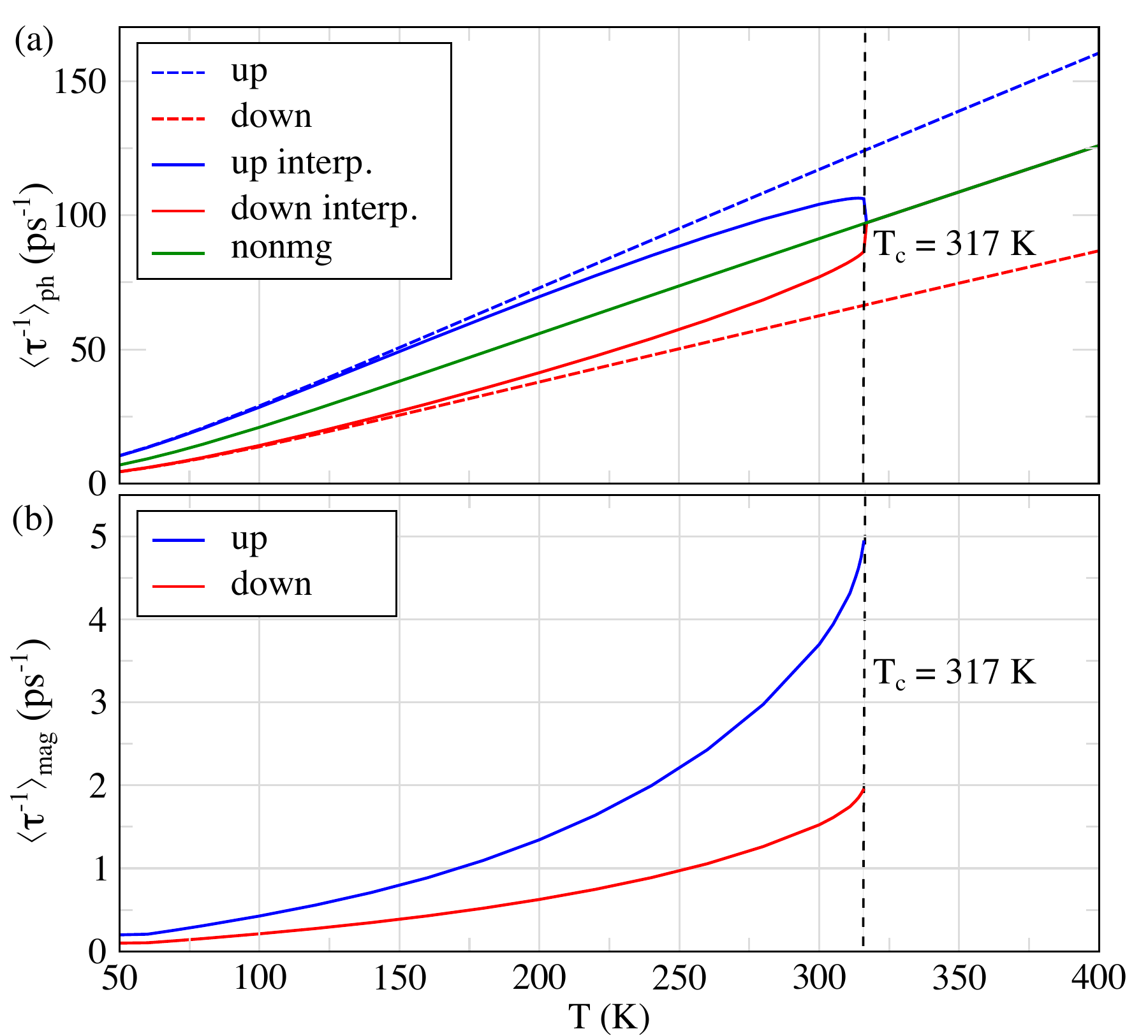}
\caption{(a) Averaged phonon-mediated scattering rate calculated for different spin channels in the ferromagnetic phase of monolayer \fgt{} (dashed line) along with the result of interpolation (solid red and blue lines), as well as in the nonmagnetic phase (green line). (b) Averaged magnon-mediated scattering rate calculated for different spin channels.}
\label{fig:Tau_T}
\end{figure}

In Fig.~\ref{fig:Tau_T}(a), we show averaged phonon-mediated scattering rate $\braket{\tau^{-1}_\sigma} = \frac{1}{N_F^\sigma} \sum_{n\mathbf{k}} \tau^{-1}_{n\mathbf{k} \sigma} \delta(\varepsilon_{n\mathbf{k} \sigma})$ as a function of  temperature. Similarly to  the coupling constant, the scattering rate calculated for the nonmagnetic phase has the value in between the spin-up and spin-down cases. At $T = 300$ K $\braket{\tau^{-1}_{\uparrow}}$  = 117 ps$^{-1}$, $\braket{\tau^{-1}_{\downarrow}}$ = 62 ps$^{-1}$, and $\braket{\tau^{-1}_{\rm nonmg}}$ = 91 ps$^{-1}$. In the relevant temperature range, all these three curves demonstrate a linear-in-temperature behavior. Indeed, at sufficiently high temperatures phonons can be considered classically with the occupation numbers $b_{\mathbf{q}\nu} \simeq k_B T/\hbar \omega_{\mathbf{q}\nu}$, ensuring a linear dependence of the electron linewidth $\mathrm{Im}\Sigma^{\sigma}_{\mathbf{q}\nu}(T)$ as well as the scattering rate $\braket{\tau_{\sigma}^{-1}}$~\cite{Rudenko2020}. However, these results do not take temperature dependence of the electronic structure into account, which is important for magnetic systems. For this purpose, we use the following interpolation between the magnetic and nonmagnetic scattering rates
\begin{equation}
 \braket{\tau^{-1}_{\sigma} (T)}  =  \braket{\tau^{-1}_{\rm nonmg}} + \left[  \braket{\tau^{-1}_\sigma} -  \braket{\tau^{-1}_{\rm nonmg}} \right] \frac{\braket{S_z}}{S}, 
 \label{eq:tau_interpolate}
\end{equation}
which allows us to make a connection between the scattering rates above and below $T_C$. The interpolation changes the linear dependence of $\braket{\tau^{-1}}$ 
at $T>100$ K, inducing a convergence of $\braket{\tau^{-1}_{\uparrow}}$ and  $\braket{\tau^{-1}_{\downarrow}}$ to the nonmagnetic solution $\braket{\tau^{-1}_{\mathrm{nonmg}}}$ at $T=T_C$ [Fig.~\ref{fig:Tau_T}(a)]. Remarkably, the averaged spin-up and spin-down scattering rates are very close the nonmagnetic scattering rate over the whole range of temperatures.

The magnon contribution to the scattering rate is shown in Fig.~\ref{fig:Tau_T}(b). The electron-magnon coupling constant is estimated to be $I \cdot N_F \simeq $ 0.01, i.e. an order of magnitude smaller compared to the electron-phonon coupling. As a consequence, the calculated $\braket{\tau^{-1}}_{\rm  mag}$ is significantly smaller than  $\braket{\tau^{-1}}_{\rm ph}$ [Fig.~\ref{fig:Tau_T}(b)]. Overall, the magnon-mediated scattering rate demonstrates an exponential increase with temperature up to $T_C$, which can be explained by the increase of the magnon population. Nevertheless, even close to $T_C$, the contribution of magnons to the scattering rate is marginal. At $T > T_C$ the long-range ferromagnetic order disappears, i.e. $\braket{S_z}=0$, and Eq.~(\ref{eq:Selfen_magnon}) is seemingly inapplicable. However, the contribution of spin excitations does not disappear in the paramagnetic phase due to the short-range spin fluctuations~\cite{Fisher1968}. Indeed, at sufficiently large $T$ we have $b_{{\bf q}\nu}\simeq k_BT/\hbar\omega_{{\bf q}\nu}$ with $\omega_{{\bf q}\nu} \sim \langle S_z\rangle$, leading to elimination of the magnetization in Eq.~(\ref{eq:Selfen_magnon}), which ensures nonzero scattering rate by static spin fluctuations. Nevertheless, as this contribution cannot be significantly larger than the contribution from magnons in the ferromagnetic phase, we ignore it from the explicit consideration.

\subsection{Spin-resolved transport}
\begin{figure}[tbp]
\centering
\includegraphics[width=0.95\linewidth]{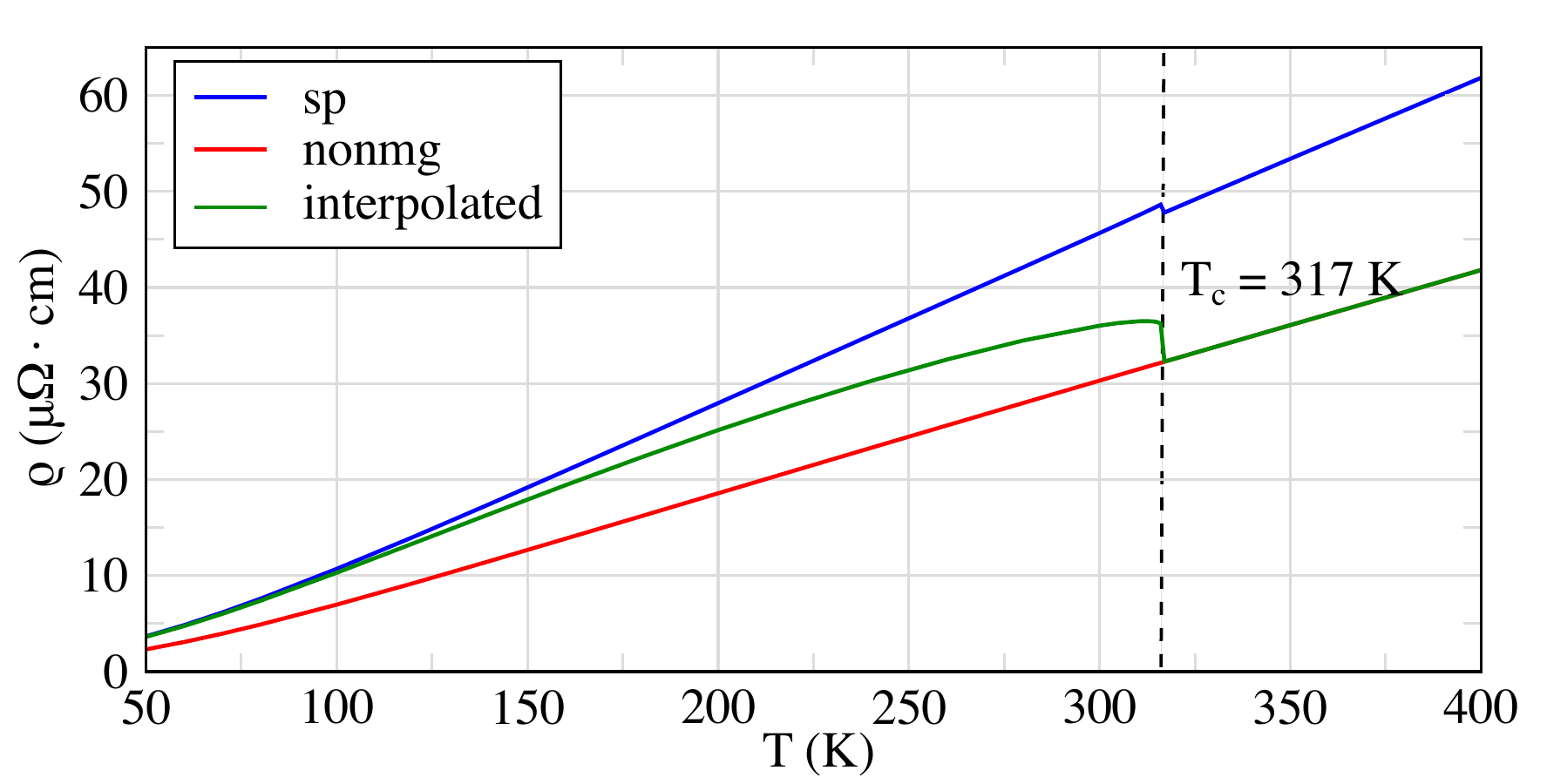}
\caption{Electrical resistivity shown as a function of temperature in monolayer \fgt{} calculated as: (i) $\rho^{-1}_{\rm sp} = \rho^{-1}_{\uparrow} + \rho^{-1}_{\downarrow}$, i.e. from fully spin-polarized calculations  without interpolation (low-$T$ limit); (ii) $\rho_{\rm  nonmg}$ i.e. nonmagnetic calculations ($T>T_C$); (iii) $\rho^{-1}(T)=\rho^{-1}_{\uparrow}(T)+\rho^{-1}_{\downarrow}(T)$   (interpolated via Eq.(\ref{interp_sigma})).}
\label{fig:Res_T}
\end{figure}

We now calculate electric resistivity $\rho = 1/ \sigma$ as a function of temperature by means of Eq.~(\ref{eq:Conductivity}). To this end, we use the Matthiessen’s rule \cite{Ziman2001_book}, $\braket{\tau^{-1}}$ =  $\braket{\tau^{-1}_{\rm ph}}$ +  $\braket{\tau^{-1}_{\rm mag}}$. Fully spin-polarized (sp)  resistivity ($\rho^{-1}_{\rm sp} =  \rho^{-1}_{\uparrow} + \rho^{-1}_{\downarrow}$) is represented in Fig.~\ref{fig:Res_T}. Due to the magnon contribution, $\rho_{\rm sp} (T)$ demonstrates a jump around $T_C$, which is attributed to the behavior of $\braket{\tau^{-1}_{\rm mag}}$ shown in Fig.~\ref{fig:Tau_T}(b).

The nonmagnetic resistivity $\rho_{\rm nonmg}$ demonstrates a less steep behavior compared to the resistivity in the ferromagnetic phase. This can be explained by the two factors: (i) temperature dependence of the scattering rate (Fig.~\ref{fig:Tau_T}); and (ii) different carrier velocities: $\overline{v}_{\uparrow, \downarrow} \simeq$  2 $\times 10^{5}$ m/s, while  $\overline{v}_{\rm nonmg}$ = 1.3  $\times 10^{5}$ m/s. The interpolation of $\overline{v} \cdot N_F$  between the magnetic and nonmagnetic phases using the procedure described above, leads to the following conductivity:
\begin{equation}
\label{interp_sigma}
    \sigma_{\sigma} (T) \simeq \frac{e^2}{\Omega} \braket{\tau_{\sigma} (T)} \braket{\overline{v}_\sigma \cdot N^{\sigma}_F (T)},   
\end{equation}
 which is shown as a green line in Fig.~\ref{fig:Res_T}. One can see that at $T>100$ K the linear resistivity behavior is modified, lowering the resistivity. At $T=T_C$ the resistivity reduces abruptly to the nonmagnetic value. Such a behavior is likely unphysical and could be attributed to our approximation of the paramagnetic phase by a nonmagnetic one. Moreover, the transition region is to be smoothed by taking into account scattering by spin inhomogeneities above the Curie temperature as well as by electron-impurity scattering. Nevertheless, our interpolated results allow us to qualitatively explain the temperature dependence of the resistivity near transition region observed in recent experiments~\cite{FGT_Transport_Feng2022, FGT_transport_Chen2013, FGT_transport_Liu2018, Deng2018, FGT_Kim2018}.

 At $T = 300$ K, the interpolated resistivity $\rho(T)$ is estimated to be 35 $\mu \Omega \cdot$cm.  The available experimental estimates for bulk FGT~\cite{FGT_Transport_Feng2022, FGT_transport_Chen2013, FGT_transport_Liu2018, Deng2018, FGT_Kim2018} vary from 150 to 200 $\mu \Omega \cdot$cm, which is 4-5 times higher than the calculated values. This disagreement can be attributed to the presence of other scattering channels in experimental samples (e.g., impurities), which also leads to  nonzero $\rho$ at $T \sim $ 0 K, as well as to the dimensionality effects. Our calculations give the lowest  boundary for the intrinsic resistivity in monolayer \fgt{}. At the  same time, this value is an order of magnitude larger than  resistivities of noble metals such  as $\rho_{\rm Cu} \sim $ 1.5 $\mu \Omega \cdot$cm and $\rho_{\rm Au} \sim $ 2.0 $\mu \Omega \cdot$cm~\cite{Jamal2016}. The relatively high resistivity of monolayer FGT might limit its prospects for electronic applications.

\section{\label{sec:Conc}Conclusions}
In summary, we have  systematically  studied the charge-carrier scattering mechanisms in ferromagnetic monolayer \fgt{}. We show that the phonon-mediated scattering of charge carriers  plays a dominant role. The effect of magnons is also present, but can be considered as negligible in the relevant temperature region. At the same time, the magnetism-induced splitting of the energy bands  modifies the electron-phonon coupling and induces a nontrivial contribution to its temperature dependence. This results in a sublinear temperature dependence  of the electrical resistivity near the ferromagnetic-paramagnetic phase transition, observed experimentally. Also, we demonstrate that the charge doping (see Appendix~\ref{app:doping}) does not lead to any pronounced changes in the transport properties of \fgt{}. Our calculations provide a lower estimate for the room temperature resistivity (35 $\mu\Omega \cdot$cm) in monolayer FGT. The resulting value is an order of magnitude smaller compared to typical metals, limiting potential applicability of monolayer FGT in electronics. The model approach presented in this paper to study transport properties of magnetic conductors can be applied to other 2D magnetic materials. Particularly, we expect a non-trivial role of the electron-magnon interactions in systems with weak electron-phonon coupling.

\section{\label{sec:Data}Data availability}
The data that support the findings of this work are available from the corresponding author upon reasonable request.

\section{\label{sec:Code}Code availability}
The central codes used in this paper are {\sc Quantum Espresso}~\cite{QE1,QE2}, {\sc EPW}~\cite{Ponce2016} and {\sc Wannier90}~\cite{pizzi2020} can be requested from developers.

\begin{acknowledgements}
The work was supported by the European Union’s Horizon 2020 research and innovation program under
European Research Council synergy grant 854843 “FASTCORR”. Calculation of exchange interactions was supported  by the Russian Science Foundation, Grant No. 21-72-10136.
\end{acknowledgements}

\appendix
\section{\label{app:ISO}Exchange interactions from first principles}
In Table~\ref{tab:Exchange_couplings},  we provide isotropic exchange interactions between iron atoms,  calculated using the local force theorem approach~\cite{liechtenstein1987, Kashin2020}:
\begin{equation}
    J_{ij} = \frac{1}{2 \pi S^2} \int \limits_{-\infty}^{E_F} d \varepsilon \,{\rm Im} \left[  \Delta_i G^\downarrow_{ij} (\varepsilon) \Delta_j G^\uparrow_{ji} (\varepsilon) \right],
\end{equation}
where  $\Delta_i$ is the on-site spin-splitting energy and $G(\varepsilon)  = (\varepsilon - \hat{H})^{-1}$ is single-particle Green's functions constructed using Hamiltonian $\hat{H}$ in the Wannier functions basis. 

\begin{table}
\centering
\caption {Main isotropic exchange interaction of \fgt{} monolayer.  Some of the pairs contain values $J^{\rm I}/J^{\rm II}$, which corresponds to different sublattices of Fe$_{\rm I}$ and Fe$_{\rm II}$ (see Fig.~\ref{fig:Structure}(a) for a crystal structure).}
\begin{ruledtabular}
\begin {tabular}{c|c|c}
 \#  &  Fe -- Fe distance (\AA)  & Exchange interaction  $J_{\#}$ (meV)  \\
 \hline 
1 &  2.493  & -115.5          \\
2 &  2.623  &  -33.5          \\
3 &  4.000  &  3.1/7.4        \\
4 &  4.712  &  2.6            \\
5 &  4.782  & -6.9            \\
6 &  6.234  &  1.8            \\ 
7 &  6.926  & -1.2/-3.3       \\
8 &  7.361  & -3.0            \\
9 &  8.000  & -3.0/0.5        \\
10& 8.377   & -0.8            \\ 
11& 8.417   &  1.2            \\
12& 9.319   &  0.1            \\
  \end {tabular}
\end{ruledtabular}
\label{tab:Exchange_couplings}
\end {table}

\section{\label{app:dlm}Density of states and disordered local moments}

\begin{figure}
\centering
\includegraphics[width=0.95\linewidth]{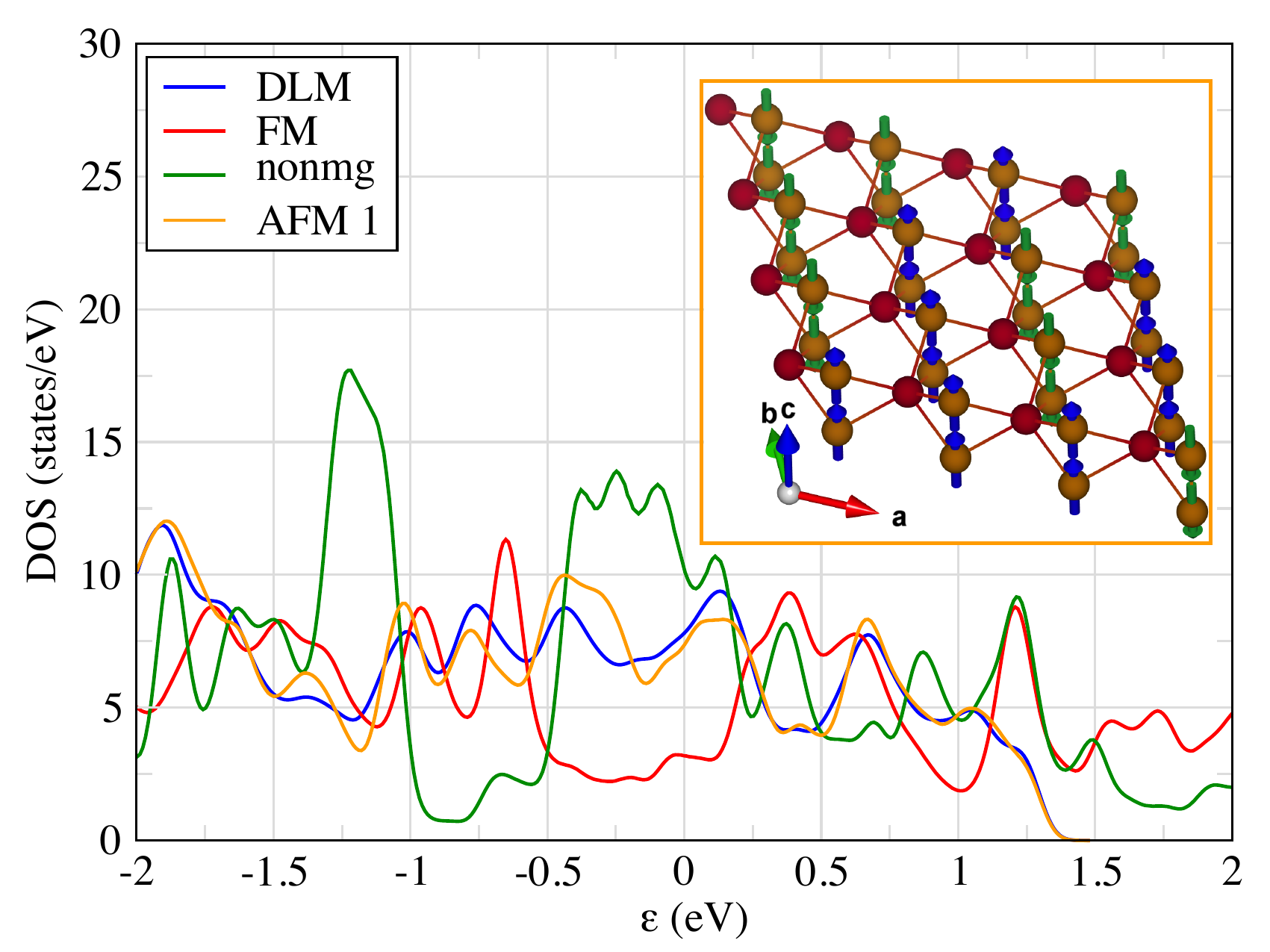}
\caption{Electron densities of states (DOS) for monolayer \fgt{} calculated in the DLM, FM and nonmagnetic state. The inset shows the magnetic configuration for one of the random AFM states used in the DLM simulations. The corresponding DOS is also shown. Zero energy corresponds to the Fermi energy.}
\label{fig:DLM}
\end{figure}

Here, we simulate the disordered local moments (DLM) state using the supercell approach described, e.g., in Ref.~\cite{Abrikosov2016}. For this purpose we consider the densities of states (DOS) averaged between $n = 20$ randomly generated collinear AFM configurations for a (4$\times$4) unit cell with zero total magnetic moment:
\begin{equation}
    N_{\rm DLM}(\varepsilon) = \frac{1}{n}\sum_{i}  N_{{\rm AFM} \,i}(\varepsilon),
\end{equation}
where $N_{{\rm AFM}\,i}(\varepsilon) =\sum_{n{\bf k}\sigma}\delta(\varepsilon^{\sigma (i)}_{n{\bf k}} - \varepsilon)$ is DOS for the random $i$th AFM state. We also check that further increase of the number of configurations does not change the result. In these sample configurations, we keep the magnetic moment of the neighboring Fe$_{\rm I}$ atoms (see Fig.~\ref{fig:Structure}(a) for a crystal structure) aligned in the same direction because they are coupled via strong FM interaction $J_1$ = $-115.5$ meV (Table~\ref{tab:Exchange_couplings}), i.e. $J_1 \gg T_C$. The calculated densities of states are shown in Fig.~\ref{fig:DLM} in comparison with the ferromagnetic and nonmagnetic solution. In addition, we show the calculated DOS for a random (AFM1) configuration, which  was used to simulate the DLM state.

\section{\label{app:doping}Role of charge doping}
\begin{figure}[t!]
\centering
\includegraphics[width=0.95\linewidth]{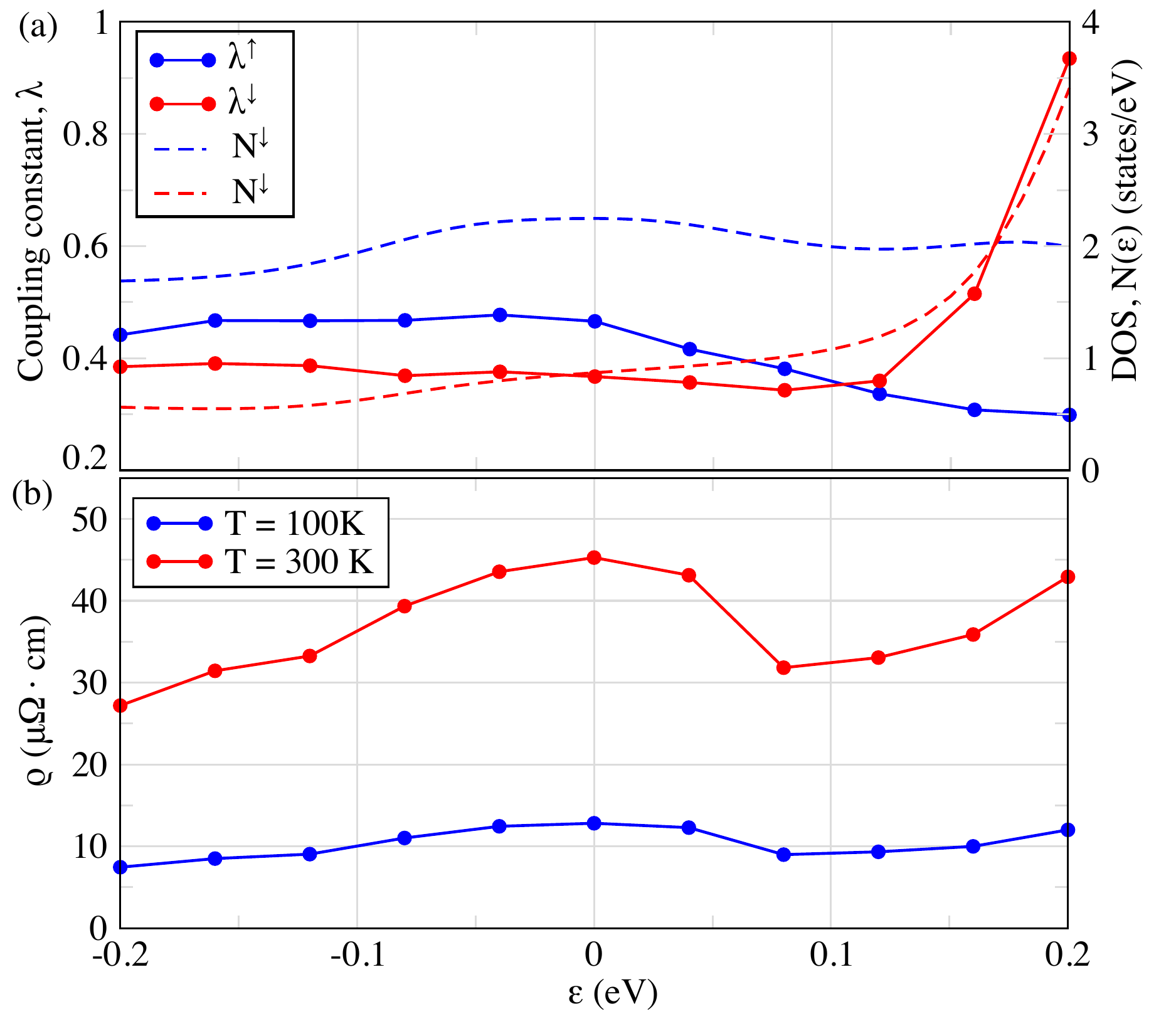}
\caption{(a) Total coupling constant $\lambda =  \sum_{\mathbf{q \nu}} \lambda_{\mathbf{q \nu}}$ and electron DOS $N^\sigma(\epsilon)$ shown as a function of the Fermi energy. (b) Resistivity for $T = 100$ K and 300 K as a function of the Fermi energy.  }
\label{fig:Doping}
\end{figure}

Here we consider the role of the Fermi energy variation on the resistivity and electron-phonon coupling constant. The rigid shift of the Fermi energy mimics the effect of charge carrier doping, and allows us to estimate to which extend the transport properties of FGT can be tuned. Both the resistivity and the coupling constant are proportional [Eqs.~(\ref{eq:Conductivity}) and (\ref{eq:Lambda})] to the electron DOS $N_F^\sigma$ at the Fermi level. Figure \ref{fig:Doping}(a) shows the electron-phonon coupling constant as well as the electron DOS as a function of the Fermi energy. One can see that $\lambda^\sigma$ is very similar to the profile of $N^\sigma (\varepsilon)$ for both spin channels. For spin-up, the Fermi energy variation within the range $-$0.2 eV -- +0.2 eV does not lead to any significant changes of $\lambda^\uparrow$. On the contrary, $\lambda^\downarrow$ demonstrates a notable enhancement at $E>0.1$ eV, such that $\lambda^\uparrow < \lambda^\downarrow$. 

The resistivity does not show any prominent dependence of the Fermi energy within the relevant energy range [Fig.~\ref{fig:Doping}(b)]. At $T = 100$ K, the resulting curve $\rho_{\rm tot}(\varepsilon)$ is essentially flat. At $T = 300$ K, the resistivity variations are more pronounced yet not significant.


\begin{thebibliography}{64}%
\makeatletter
\providecommand \@ifxundefined [1]{%
 \@ifx{#1\undefined}
}%
\providecommand \@ifnum [1]{%
 \ifnum #1\expandafter \@firstoftwo
 \else \expandafter \@secondoftwo
 \fi
}%
\providecommand \@ifx [1]{%
 \ifx #1\expandafter \@firstoftwo
 \else \expandafter \@secondoftwo
 \fi
}%
\providecommand \natexlab [1]{#1}%
\providecommand \enquote  [1]{``#1''}%
\providecommand \bibnamefont  [1]{#1}%
\providecommand \bibfnamefont [1]{#1}%
\providecommand \citenamefont [1]{#1}%
\providecommand \href@noop [0]{\@secondoftwo}%
\providecommand \href [0]{\begingroup \@sanitize@url \@href}%
\providecommand \@href[1]{\@@startlink{#1}\@@href}%
\providecommand \@@href[1]{\endgroup#1\@@endlink}%
\providecommand \@sanitize@url [0]{\catcode `\\12\catcode `\$12\catcode
  `\&12\catcode `\#12\catcode `\^12\catcode `\_12\catcode `\%12\relax}%
\providecommand \@@startlink[1]{}%
\providecommand \@@endlink[0]{}%
\providecommand \url  [0]{\begingroup\@sanitize@url \@url }%
\providecommand \@url [1]{\endgroup\@href {#1}{\urlprefix }}%
\providecommand \urlprefix  [0]{URL }%
\providecommand \Eprint [0]{\href }%
\providecommand \doibase [0]{http://dx.doi.org/}%
\providecommand \selectlanguage [0]{\@gobble}%
\providecommand \bibinfo  [0]{\@secondoftwo}%
\providecommand \bibfield  [0]{\@secondoftwo}%
\providecommand \translation [1]{[#1]}%
\providecommand \BibitemOpen [0]{}%
\providecommand \bibitemStop [0]{}%
\providecommand \bibitemNoStop [0]{.\EOS\space}%
\providecommand \EOS [0]{\spacefactor3000\relax}%
\providecommand \BibitemShut  [1]{\csname bibitem#1\endcsname}%
\let\auto@bib@innerbib\@empty
\bibitem [{\citenamefont {Huang}\ \emph {et~al.}(2017)\citenamefont {Huang},
  \citenamefont {Clark}, \citenamefont {Navarro-Moratalla}, \citenamefont
  {Klein}, \citenamefont {Cheng}, \citenamefont {Seyler}, \citenamefont
  {Zhong}, \citenamefont {Schmidgall}, \citenamefont {McGuire}, \citenamefont
  {Cobden}, \citenamefont {Yao}, \citenamefont {Xiao}, \citenamefont
  {Jarillo-Herrero},\ and\ \citenamefont {Xu}}]{CrI3_Huang2017}%
  \BibitemOpen
  \bibfield  {author} {\bibinfo {author} {\bibfnamefont {B.}~\bibnamefont
  {Huang}}, \bibinfo {author} {\bibfnamefont {G.}~\bibnamefont {Clark}},
  \bibinfo {author} {\bibfnamefont {E.}~\bibnamefont {Navarro-Moratalla}},
  \bibinfo {author} {\bibfnamefont {D.~R.}\ \bibnamefont {Klein}}, \bibinfo
  {author} {\bibfnamefont {R.}~\bibnamefont {Cheng}}, \bibinfo {author}
  {\bibfnamefont {K.~L.}\ \bibnamefont {Seyler}}, \bibinfo {author}
  {\bibfnamefont {D.}~\bibnamefont {Zhong}}, \bibinfo {author} {\bibfnamefont
  {E.}~\bibnamefont {Schmidgall}}, \bibinfo {author} {\bibfnamefont {M.~A.}\
  \bibnamefont {McGuire}}, \bibinfo {author} {\bibfnamefont {D.~H.}\
  \bibnamefont {Cobden}}, \bibinfo {author} {\bibfnamefont {W.}~\bibnamefont
  {Yao}}, \bibinfo {author} {\bibfnamefont {D.}~\bibnamefont {Xiao}}, \bibinfo
  {author} {\bibfnamefont {P.}~\bibnamefont {Jarillo-Herrero}}, \ and\ \bibinfo
  {author} {\bibfnamefont {X.}~\bibnamefont {Xu}},\ }\href {\doibase
  10.1038/nature22391} {\bibfield  {journal} {\bibinfo  {journal} {Nature}\
  }\textbf {\bibinfo {volume} {546}},\ \bibinfo {pages} {270} (\bibinfo {year}
  {2017})}\BibitemShut {NoStop}%
\bibitem [{\citenamefont {Gong}\ \emph {et~al.}(2017)\citenamefont {Gong},
  \citenamefont {Li}, \citenamefont {Li}, \citenamefont {Ji}, \citenamefont
  {Stern}, \citenamefont {Xia}, \citenamefont {Cao}, \citenamefont {Bao},
  \citenamefont {Wang}, \citenamefont {Wang}, \citenamefont {Qiu},
  \citenamefont {Cava}, \citenamefont {Louie}, \citenamefont {Xia},\ and\
  \citenamefont {Zhang}}]{Cr2Ge2Te6_Gong2017}%
  \BibitemOpen
  \bibfield  {author} {\bibinfo {author} {\bibfnamefont {C.}~\bibnamefont
  {Gong}}, \bibinfo {author} {\bibfnamefont {L.}~\bibnamefont {Li}}, \bibinfo
  {author} {\bibfnamefont {Z.}~\bibnamefont {Li}}, \bibinfo {author}
  {\bibfnamefont {H.}~\bibnamefont {Ji}}, \bibinfo {author} {\bibfnamefont
  {A.}~\bibnamefont {Stern}}, \bibinfo {author} {\bibfnamefont
  {Y.}~\bibnamefont {Xia}}, \bibinfo {author} {\bibfnamefont {T.}~\bibnamefont
  {Cao}}, \bibinfo {author} {\bibfnamefont {W.}~\bibnamefont {Bao}}, \bibinfo
  {author} {\bibfnamefont {C.}~\bibnamefont {Wang}}, \bibinfo {author}
  {\bibfnamefont {Y.}~\bibnamefont {Wang}}, \bibinfo {author} {\bibfnamefont
  {Z.~Q.}\ \bibnamefont {Qiu}}, \bibinfo {author} {\bibfnamefont {R.~J.}\
  \bibnamefont {Cava}}, \bibinfo {author} {\bibfnamefont {S.~G.}\ \bibnamefont
  {Louie}}, \bibinfo {author} {\bibfnamefont {J.}~\bibnamefont {Xia}}, \ and\
  \bibinfo {author} {\bibfnamefont {X.}~\bibnamefont {Zhang}},\ }\href
  {\doibase 10.1038/nature22060} {\bibfield  {journal} {\bibinfo  {journal}
  {Nature}\ }\textbf {\bibinfo {volume} {546}},\ \bibinfo {pages} {265}
  (\bibinfo {year} {2017})}\BibitemShut {NoStop}%
\bibitem [{\citenamefont {Deiseroth}\ \emph
  {et~al.}(2006{\natexlab{a}})\citenamefont {Deiseroth}, \citenamefont
  {Aleksandrov}, \citenamefont {Reiner}, \citenamefont {Kienle},\ and\
  \citenamefont {Kremer}}]{FGT_Deiseroth2006}%
  \BibitemOpen
  \bibfield  {author} {\bibinfo {author} {\bibfnamefont {H.-J.}\ \bibnamefont
  {Deiseroth}}, \bibinfo {author} {\bibfnamefont {K.}~\bibnamefont
  {Aleksandrov}}, \bibinfo {author} {\bibfnamefont {C.}~\bibnamefont {Reiner}},
  \bibinfo {author} {\bibfnamefont {L.}~\bibnamefont {Kienle}}, \ and\ \bibinfo
  {author} {\bibfnamefont {R.~K.}\ \bibnamefont {Kremer}},\ }\href {\doibase
  https://doi.org/10.1002/ejic.200501020} {\bibfield  {journal} {\bibinfo
  {journal} {European Journal of Inorganic Chemistry}\ }\textbf {\bibinfo
  {volume} {2006}},\ \bibinfo {pages} {1561} (\bibinfo {year}
  {2006}{\natexlab{a}})}\BibitemShut {NoStop}%
\bibitem [{\citenamefont {Deng}\ \emph
  {et~al.}(2018{\natexlab{a}})\citenamefont {Deng}, \citenamefont {Yu},
  \citenamefont {Song}, \citenamefont {Zhang}, \citenamefont {Wang},
  \citenamefont {Sun}, \citenamefont {Yi}, \citenamefont {Wu}, \citenamefont
  {Wu}, \citenamefont {Zhu}, \citenamefont {Wang}, \citenamefont {Chen},\ and\
  \citenamefont {Zhang}}]{FGT_Deng2018}%
  \BibitemOpen
  \bibfield  {author} {\bibinfo {author} {\bibfnamefont {Y.}~\bibnamefont
  {Deng}}, \bibinfo {author} {\bibfnamefont {Y.}~\bibnamefont {Yu}}, \bibinfo
  {author} {\bibfnamefont {Y.}~\bibnamefont {Song}}, \bibinfo {author}
  {\bibfnamefont {J.}~\bibnamefont {Zhang}}, \bibinfo {author} {\bibfnamefont
  {N.~Z.}\ \bibnamefont {Wang}}, \bibinfo {author} {\bibfnamefont
  {Z.}~\bibnamefont {Sun}}, \bibinfo {author} {\bibfnamefont {Y.}~\bibnamefont
  {Yi}}, \bibinfo {author} {\bibfnamefont {Y.~Z.}\ \bibnamefont {Wu}}, \bibinfo
  {author} {\bibfnamefont {S.}~\bibnamefont {Wu}}, \bibinfo {author}
  {\bibfnamefont {J.}~\bibnamefont {Zhu}}, \bibinfo {author} {\bibfnamefont
  {J.}~\bibnamefont {Wang}}, \bibinfo {author} {\bibfnamefont {X.~H.}\
  \bibnamefont {Chen}}, \ and\ \bibinfo {author} {\bibfnamefont
  {Y.}~\bibnamefont {Zhang}},\ }\href {\doibase 10.1038/s41586-018-0626-9}
  {\bibfield  {journal} {\bibinfo  {journal} {Nature}\ }\textbf {\bibinfo
  {volume} {563}},\ \bibinfo {pages} {94} (\bibinfo {year}
  {2018}{\natexlab{a}})}\BibitemShut {NoStop}%
\bibitem [{\citenamefont {Fei}\ \emph {et~al.}(2018{\natexlab{a}})\citenamefont
  {Fei}, \citenamefont {Huang}, \citenamefont {Malinowski}, \citenamefont
  {Wang}, \citenamefont {Song}, \citenamefont {Sanchez}, \citenamefont {Yao},
  \citenamefont {Xiao}, \citenamefont {Zhu}, \citenamefont {May}, \citenamefont
  {Wu}, \citenamefont {Cobden}, \citenamefont {Chu},\ and\ \citenamefont
  {Xu}}]{FGT_Fei2018}%
  \BibitemOpen
  \bibfield  {author} {\bibinfo {author} {\bibfnamefont {Z.}~\bibnamefont
  {Fei}}, \bibinfo {author} {\bibfnamefont {B.}~\bibnamefont {Huang}}, \bibinfo
  {author} {\bibfnamefont {P.}~\bibnamefont {Malinowski}}, \bibinfo {author}
  {\bibfnamefont {W.}~\bibnamefont {Wang}}, \bibinfo {author} {\bibfnamefont
  {T.}~\bibnamefont {Song}}, \bibinfo {author} {\bibfnamefont {J.}~\bibnamefont
  {Sanchez}}, \bibinfo {author} {\bibfnamefont {W.}~\bibnamefont {Yao}},
  \bibinfo {author} {\bibfnamefont {D.}~\bibnamefont {Xiao}}, \bibinfo {author}
  {\bibfnamefont {X.}~\bibnamefont {Zhu}}, \bibinfo {author} {\bibfnamefont
  {A.~F.}\ \bibnamefont {May}}, \bibinfo {author} {\bibfnamefont
  {W.}~\bibnamefont {Wu}}, \bibinfo {author} {\bibfnamefont {D.~H.}\
  \bibnamefont {Cobden}}, \bibinfo {author} {\bibfnamefont {J.-H.}\
  \bibnamefont {Chu}}, \ and\ \bibinfo {author} {\bibfnamefont
  {X.}~\bibnamefont {Xu}},\ }\href {\doibase 10.1038/s41563-018-0149-7}
  {\bibfield  {journal} {\bibinfo  {journal} {Nature Materials}\ }\textbf
  {\bibinfo {volume} {17}},\ \bibinfo {pages} {778} (\bibinfo {year}
  {2018}{\natexlab{a}})}\BibitemShut {NoStop}%
\bibitem [{\citenamefont {Li}\ \emph {et~al.}(2018)\citenamefont {Li},
  \citenamefont {Yang}, \citenamefont {Gong}, \citenamefont {Chopdekar},
  \citenamefont {N’Diaye}, \citenamefont {Turner}, \citenamefont {Chen},
  \citenamefont {Scholl}, \citenamefont {Shafer}, \citenamefont {Arenholz},
  \citenamefont {Schmid}, \citenamefont {Wang}, \citenamefont {Liu},
  \citenamefont {Gao}, \citenamefont {Admasu}, \citenamefont {Cheong},
  \citenamefont {Hwang}, \citenamefont {Li}, \citenamefont {Wang},
  \citenamefont {Zhang},\ and\ \citenamefont {Qiu}}]{FGT_Li2018}%
  \BibitemOpen
  \bibfield  {author} {\bibinfo {author} {\bibfnamefont {Q.}~\bibnamefont
  {Li}}, \bibinfo {author} {\bibfnamefont {M.}~\bibnamefont {Yang}}, \bibinfo
  {author} {\bibfnamefont {C.}~\bibnamefont {Gong}}, \bibinfo {author}
  {\bibfnamefont {R.~V.}\ \bibnamefont {Chopdekar}}, \bibinfo {author}
  {\bibfnamefont {A.~T.}\ \bibnamefont {N’Diaye}}, \bibinfo {author}
  {\bibfnamefont {J.}~\bibnamefont {Turner}}, \bibinfo {author} {\bibfnamefont
  {G.}~\bibnamefont {Chen}}, \bibinfo {author} {\bibfnamefont {A.}~\bibnamefont
  {Scholl}}, \bibinfo {author} {\bibfnamefont {P.}~\bibnamefont {Shafer}},
  \bibinfo {author} {\bibfnamefont {E.}~\bibnamefont {Arenholz}}, \bibinfo
  {author} {\bibfnamefont {A.~K.}\ \bibnamefont {Schmid}}, \bibinfo {author}
  {\bibfnamefont {S.}~\bibnamefont {Wang}}, \bibinfo {author} {\bibfnamefont
  {K.}~\bibnamefont {Liu}}, \bibinfo {author} {\bibfnamefont {N.}~\bibnamefont
  {Gao}}, \bibinfo {author} {\bibfnamefont {A.~S.}\ \bibnamefont {Admasu}},
  \bibinfo {author} {\bibfnamefont {S.-W.}\ \bibnamefont {Cheong}}, \bibinfo
  {author} {\bibfnamefont {C.}~\bibnamefont {Hwang}}, \bibinfo {author}
  {\bibfnamefont {J.}~\bibnamefont {Li}}, \bibinfo {author} {\bibfnamefont
  {F.}~\bibnamefont {Wang}}, \bibinfo {author} {\bibfnamefont {X.}~\bibnamefont
  {Zhang}}, \ and\ \bibinfo {author} {\bibfnamefont {Z.}~\bibnamefont {Qiu}},\
  }\href {\doibase 10.1021/acs.nanolett.8b02806} {\bibfield  {journal}
  {\bibinfo  {journal} {Nano Letters}\ }\textbf {\bibinfo {volume} {18}},\
  \bibinfo {pages} {5974} (\bibinfo {year} {2018})}\BibitemShut {NoStop}%
\bibitem [{\citenamefont {Huang}\ \emph {et~al.}(2018)\citenamefont {Huang},
  \citenamefont {Clark}, \citenamefont {Klein}, \citenamefont {MacNeill},
  \citenamefont {Navarro-Moratalla}, \citenamefont {Seyler}, \citenamefont
  {Wilson}, \citenamefont {McGuire}, \citenamefont {Cobden}, \citenamefont
  {Xiao}, \citenamefont {Yao}, \citenamefont {Jarillo-Herrero},\ and\
  \citenamefont {Xu}}]{Huang2018}%
  \BibitemOpen
  \bibfield  {author} {\bibinfo {author} {\bibfnamefont {B.}~\bibnamefont
  {Huang}}, \bibinfo {author} {\bibfnamefont {G.}~\bibnamefont {Clark}},
  \bibinfo {author} {\bibfnamefont {D.~R.}\ \bibnamefont {Klein}}, \bibinfo
  {author} {\bibfnamefont {D.}~\bibnamefont {MacNeill}}, \bibinfo {author}
  {\bibfnamefont {E.}~\bibnamefont {Navarro-Moratalla}}, \bibinfo {author}
  {\bibfnamefont {K.~L.}\ \bibnamefont {Seyler}}, \bibinfo {author}
  {\bibfnamefont {N.}~\bibnamefont {Wilson}}, \bibinfo {author} {\bibfnamefont
  {M.~A.}\ \bibnamefont {McGuire}}, \bibinfo {author} {\bibfnamefont {D.~H.}\
  \bibnamefont {Cobden}}, \bibinfo {author} {\bibfnamefont {D.}~\bibnamefont
  {Xiao}}, \bibinfo {author} {\bibfnamefont {W.}~\bibnamefont {Yao}}, \bibinfo
  {author} {\bibfnamefont {P.}~\bibnamefont {Jarillo-Herrero}}, \ and\ \bibinfo
  {author} {\bibfnamefont {X.}~\bibnamefont {Xu}},\ }\href {\doibase
  10.1038/s41565-018-0121-3} {\bibfield  {journal} {\bibinfo  {journal} {Nature
  Nanotechnology}\ }\textbf {\bibinfo {volume} {13}},\ \bibinfo {pages} {544}
  (\bibinfo {year} {2018})}\BibitemShut {NoStop}%
\bibitem [{\citenamefont {Soriano}\ \emph {et~al.}(2021)\citenamefont
  {Soriano}, \citenamefont {Rudenko}, \citenamefont {Katsnelson},\ and\
  \citenamefont {R{\"o}sner}}]{Soriano2021}%
  \BibitemOpen
  \bibfield  {author} {\bibinfo {author} {\bibfnamefont {D.}~\bibnamefont
  {Soriano}}, \bibinfo {author} {\bibfnamefont {A.~N.}\ \bibnamefont
  {Rudenko}}, \bibinfo {author} {\bibfnamefont {M.~I.}\ \bibnamefont
  {Katsnelson}}, \ and\ \bibinfo {author} {\bibfnamefont {M.}~\bibnamefont
  {R{\"o}sner}},\ }\href {\doibase 10.1038/s41524-021-00631-4} {\bibfield
  {journal} {\bibinfo  {journal} {npj Computational Materials}\ }\textbf
  {\bibinfo {volume} {7}},\ \bibinfo {pages} {162} (\bibinfo {year}
  {2021})}\BibitemShut {NoStop}%
\bibitem [{\citenamefont {Kim}\ \emph {et~al.}(2018)\citenamefont {Kim},
  \citenamefont {Seo}, \citenamefont {Lee}, \citenamefont {Ko}, \citenamefont
  {Kim}, \citenamefont {Jang}, \citenamefont {Ok}, \citenamefont {Lee},
  \citenamefont {Jo}, \citenamefont {Kang}, \citenamefont {Shim}, \citenamefont
  {Kim}, \citenamefont {Yeom}, \citenamefont {Il~Min}, \citenamefont {Yang},\
  and\ \citenamefont {Kim}}]{FGT_Kim2018}%
  \BibitemOpen
  \bibfield  {author} {\bibinfo {author} {\bibfnamefont {K.}~\bibnamefont
  {Kim}}, \bibinfo {author} {\bibfnamefont {J.}~\bibnamefont {Seo}}, \bibinfo
  {author} {\bibfnamefont {E.}~\bibnamefont {Lee}}, \bibinfo {author}
  {\bibfnamefont {K.-T.}\ \bibnamefont {Ko}}, \bibinfo {author} {\bibfnamefont
  {B.~S.}\ \bibnamefont {Kim}}, \bibinfo {author} {\bibfnamefont {B.~G.}\
  \bibnamefont {Jang}}, \bibinfo {author} {\bibfnamefont {J.~M.}\ \bibnamefont
  {Ok}}, \bibinfo {author} {\bibfnamefont {J.}~\bibnamefont {Lee}}, \bibinfo
  {author} {\bibfnamefont {Y.~J.}\ \bibnamefont {Jo}}, \bibinfo {author}
  {\bibfnamefont {W.}~\bibnamefont {Kang}}, \bibinfo {author} {\bibfnamefont
  {J.~H.}\ \bibnamefont {Shim}}, \bibinfo {author} {\bibfnamefont
  {C.}~\bibnamefont {Kim}}, \bibinfo {author} {\bibfnamefont {H.~W.}\
  \bibnamefont {Yeom}}, \bibinfo {author} {\bibfnamefont {B.}~\bibnamefont
  {Il~Min}}, \bibinfo {author} {\bibfnamefont {B.-J.}\ \bibnamefont {Yang}}, \
  and\ \bibinfo {author} {\bibfnamefont {J.~S.}\ \bibnamefont {Kim}},\ }\href
  {\doibase 10.1038/s41563-018-0132-3} {\bibfield  {journal} {\bibinfo
  {journal} {Nature Materials}\ }\textbf {\bibinfo {volume} {17}},\ \bibinfo
  {pages} {794} (\bibinfo {year} {2018})}\BibitemShut {NoStop}%
\bibitem [{\citenamefont {Roemer}\ \emph {et~al.}(2020)\citenamefont {Roemer},
  \citenamefont {Liu},\ and\ \citenamefont {Zou}}]{FGT_Transport_Roemer2020}%
  \BibitemOpen
  \bibfield  {author} {\bibinfo {author} {\bibfnamefont {R.}~\bibnamefont
  {Roemer}}, \bibinfo {author} {\bibfnamefont {C.}~\bibnamefont {Liu}}, \ and\
  \bibinfo {author} {\bibfnamefont {K.}~\bibnamefont {Zou}},\ }\href {\doibase
  10.1038/s41699-020-00167-z} {\bibfield  {journal} {\bibinfo  {journal} {npj
  2D Materials and Applications}\ }\textbf {\bibinfo {volume} {4}},\ \bibinfo
  {pages} {33} (\bibinfo {year} {2020})}\BibitemShut {NoStop}%
\bibitem [{\citenamefont {Chen}\ \emph
  {et~al.}(2013{\natexlab{a}})\citenamefont {Chen}, \citenamefont {Yang},
  \citenamefont {Wang}, \citenamefont {Imai}, \citenamefont {Ohta},
  \citenamefont {Michioka}, \citenamefont {Yoshimura},\ and\ \citenamefont
  {Fang}}]{FGT_Chen2013}%
  \BibitemOpen
  \bibfield  {author} {\bibinfo {author} {\bibfnamefont {B.}~\bibnamefont
  {Chen}}, \bibinfo {author} {\bibfnamefont {J.}~\bibnamefont {Yang}}, \bibinfo
  {author} {\bibfnamefont {H.}~\bibnamefont {Wang}}, \bibinfo {author}
  {\bibfnamefont {M.}~\bibnamefont {Imai}}, \bibinfo {author} {\bibfnamefont
  {H.}~\bibnamefont {Ohta}}, \bibinfo {author} {\bibfnamefont {C.}~\bibnamefont
  {Michioka}}, \bibinfo {author} {\bibfnamefont {K.}~\bibnamefont {Yoshimura}},
  \ and\ \bibinfo {author} {\bibfnamefont {M.}~\bibnamefont {Fang}},\ }\href
  {\doibase 10.7566/JPSJ.82.124711} {\bibfield  {journal} {\bibinfo  {journal}
  {Journal of the Physical Society of Japan}\ }\textbf {\bibinfo {volume}
  {82}},\ \bibinfo {pages} {124711} (\bibinfo {year}
  {2013}{\natexlab{a}})}\BibitemShut {NoStop}%
\bibitem [{\citenamefont {Laref}\ \emph {et~al.}(2020)\citenamefont {Laref},
  \citenamefont {Kim},\ and\ \citenamefont {Manchon}}]{FGT_DMI_Laref2020}%
  \BibitemOpen
  \bibfield  {author} {\bibinfo {author} {\bibfnamefont {S.}~\bibnamefont
  {Laref}}, \bibinfo {author} {\bibfnamefont {K.-W.}\ \bibnamefont {Kim}}, \
  and\ \bibinfo {author} {\bibfnamefont {A.}~\bibnamefont {Manchon}},\ }\href
  {\doibase 10.1103/PhysRevB.102.060402} {\bibfield  {journal} {\bibinfo
  {journal} {Phys. Rev. B}\ }\textbf {\bibinfo {volume} {102}},\ \bibinfo
  {pages} {060402} (\bibinfo {year} {2020})}\BibitemShut {NoStop}%
\bibitem [{\citenamefont {Ado}\ \emph {et~al.}(2022)\citenamefont {Ado},
  \citenamefont {Rakhmanova}, \citenamefont {Zezyulin}, \citenamefont {Iorsh},\
  and\ \citenamefont {Titov}}]{FGT_Ado2021}%
  \BibitemOpen
  \bibfield  {author} {\bibinfo {author} {\bibfnamefont {I.~A.}\ \bibnamefont
  {Ado}}, \bibinfo {author} {\bibfnamefont {G.}~\bibnamefont {Rakhmanova}},
  \bibinfo {author} {\bibfnamefont {D.~A.}\ \bibnamefont {Zezyulin}}, \bibinfo
  {author} {\bibfnamefont {I.}~\bibnamefont {Iorsh}}, \ and\ \bibinfo {author}
  {\bibfnamefont {M.}~\bibnamefont {Titov}},\ }\href {\doibase
  10.1103/PhysRevB.106.144407} {\bibfield  {journal} {\bibinfo  {journal}
  {Phys. Rev. B}\ }\textbf {\bibinfo {volume} {106}},\ \bibinfo {pages}
  {144407} (\bibinfo {year} {2022})}\BibitemShut {NoStop}%
\bibitem [{\citenamefont {Ding}\ \emph {et~al.}(2020)\citenamefont {Ding},
  \citenamefont {Li}, \citenamefont {Xu}, \citenamefont {Li}, \citenamefont
  {Hou}, \citenamefont {Liu}, \citenamefont {Xi}, \citenamefont {Xu},
  \citenamefont {Yao},\ and\ \citenamefont {Wang}}]{FGT_Ding2020}%
  \BibitemOpen
  \bibfield  {author} {\bibinfo {author} {\bibfnamefont {B.}~\bibnamefont
  {Ding}}, \bibinfo {author} {\bibfnamefont {Z.}~\bibnamefont {Li}}, \bibinfo
  {author} {\bibfnamefont {G.}~\bibnamefont {Xu}}, \bibinfo {author}
  {\bibfnamefont {H.}~\bibnamefont {Li}}, \bibinfo {author} {\bibfnamefont
  {Z.}~\bibnamefont {Hou}}, \bibinfo {author} {\bibfnamefont {E.}~\bibnamefont
  {Liu}}, \bibinfo {author} {\bibfnamefont {X.}~\bibnamefont {Xi}}, \bibinfo
  {author} {\bibfnamefont {F.}~\bibnamefont {Xu}}, \bibinfo {author}
  {\bibfnamefont {Y.}~\bibnamefont {Yao}}, \ and\ \bibinfo {author}
  {\bibfnamefont {W.}~\bibnamefont {Wang}},\ }\href {\doibase
  10.1021/acs.nanolett.9b03453} {\bibfield  {journal} {\bibinfo  {journal}
  {Nano Letters}\ }\textbf {\bibinfo {volume} {20}},\ \bibinfo {pages} {868}
  (\bibinfo {year} {2020})}\BibitemShut {NoStop}%
\bibitem [{\citenamefont {Meijer}\ \emph {et~al.}(2020)\citenamefont {Meijer},
  \citenamefont {Lucassen}, \citenamefont {Duine}, \citenamefont {Swagten},
  \citenamefont {Koopmans}, \citenamefont {Lavrijsen},\ and\ \citenamefont
  {Guimaraes}}]{FGT_skyrmions_Meijer2020}%
  \BibitemOpen
  \bibfield  {author} {\bibinfo {author} {\bibfnamefont {M.~J.}\ \bibnamefont
  {Meijer}}, \bibinfo {author} {\bibfnamefont {J.}~\bibnamefont {Lucassen}},
  \bibinfo {author} {\bibfnamefont {R.~A.}\ \bibnamefont {Duine}}, \bibinfo
  {author} {\bibfnamefont {H.~J.}\ \bibnamefont {Swagten}}, \bibinfo {author}
  {\bibfnamefont {B.}~\bibnamefont {Koopmans}}, \bibinfo {author}
  {\bibfnamefont {R.}~\bibnamefont {Lavrijsen}}, \ and\ \bibinfo {author}
  {\bibfnamefont {M.~H.~D.}\ \bibnamefont {Guimaraes}},\ }\href {\doibase
  10.1021/acs.nanolett.0c03111} {\bibfield  {journal} {\bibinfo  {journal}
  {Nano Letters}\ }\textbf {\bibinfo {volume} {20}},\ \bibinfo {pages} {8563}
  (\bibinfo {year} {2020})}\BibitemShut {NoStop}%
\bibitem [{\citenamefont {Park}\ \emph {et~al.}(2021)\citenamefont {Park},
  \citenamefont {Peng}, \citenamefont {Liang}, \citenamefont {Hallal},
  \citenamefont {Yasin}, \citenamefont {Zhang}, \citenamefont {Song},
  \citenamefont {Kim}, \citenamefont {Kim}, \citenamefont {Weigand},
  \citenamefont {Sch\"utz}, \citenamefont {Finizio}, \citenamefont {Raabe},
  \citenamefont {Garcia}, \citenamefont {Xia}, \citenamefont {Zhou},
  \citenamefont {Ezawa}, \citenamefont {Liu}, \citenamefont {Chang},
  \citenamefont {Koo}, \citenamefont {Kim}, \citenamefont {Chshiev},
  \citenamefont {Fert}, \citenamefont {Yang}, \citenamefont {Yu},\ and\
  \citenamefont {Woo}}]{FGT_skyrmions_Park2021}%
  \BibitemOpen
  \bibfield  {author} {\bibinfo {author} {\bibfnamefont {T.-E.}\ \bibnamefont
  {Park}}, \bibinfo {author} {\bibfnamefont {L.}~\bibnamefont {Peng}}, \bibinfo
  {author} {\bibfnamefont {J.}~\bibnamefont {Liang}}, \bibinfo {author}
  {\bibfnamefont {A.}~\bibnamefont {Hallal}}, \bibinfo {author} {\bibfnamefont
  {F.~S.}\ \bibnamefont {Yasin}}, \bibinfo {author} {\bibfnamefont
  {X.}~\bibnamefont {Zhang}}, \bibinfo {author} {\bibfnamefont {K.~M.}\
  \bibnamefont {Song}}, \bibinfo {author} {\bibfnamefont {S.~J.}\ \bibnamefont
  {Kim}}, \bibinfo {author} {\bibfnamefont {K.}~\bibnamefont {Kim}}, \bibinfo
  {author} {\bibfnamefont {M.}~\bibnamefont {Weigand}}, \bibinfo {author}
  {\bibfnamefont {G.}~\bibnamefont {Sch\"utz}}, \bibinfo {author}
  {\bibfnamefont {S.}~\bibnamefont {Finizio}}, \bibinfo {author} {\bibfnamefont
  {J.}~\bibnamefont {Raabe}}, \bibinfo {author} {\bibfnamefont
  {K.}~\bibnamefont {Garcia}}, \bibinfo {author} {\bibfnamefont
  {J.}~\bibnamefont {Xia}}, \bibinfo {author} {\bibfnamefont {Y.}~\bibnamefont
  {Zhou}}, \bibinfo {author} {\bibfnamefont {M.}~\bibnamefont {Ezawa}},
  \bibinfo {author} {\bibfnamefont {X.}~\bibnamefont {Liu}}, \bibinfo {author}
  {\bibfnamefont {J.}~\bibnamefont {Chang}}, \bibinfo {author} {\bibfnamefont
  {H.~C.}\ \bibnamefont {Koo}}, \bibinfo {author} {\bibfnamefont {Y.~D.}\
  \bibnamefont {Kim}}, \bibinfo {author} {\bibfnamefont {M.}~\bibnamefont
  {Chshiev}}, \bibinfo {author} {\bibfnamefont {A.}~\bibnamefont {Fert}},
  \bibinfo {author} {\bibfnamefont {H.}~\bibnamefont {Yang}}, \bibinfo {author}
  {\bibfnamefont {X.}~\bibnamefont {Yu}}, \ and\ \bibinfo {author}
  {\bibfnamefont {S.}~\bibnamefont {Woo}},\ }\href {\doibase
  10.1103/PhysRevB.103.104410} {\bibfield  {journal} {\bibinfo  {journal}
  {Phys. Rev. B}\ }\textbf {\bibinfo {volume} {103}},\ \bibinfo {pages}
  {104410} (\bibinfo {year} {2021})}\BibitemShut {NoStop}%
\bibitem [{\citenamefont {M\"uller}\ \emph {et~al.}(2019)\citenamefont
  {M\"uller}, \citenamefont {Bl\"ugel},\ and\ \citenamefont
  {Friedrich}}]{Friedrich2019}%
  \BibitemOpen
  \bibfield  {author} {\bibinfo {author} {\bibfnamefont {M.~C. T.~D.}\
  \bibnamefont {M\"uller}}, \bibinfo {author} {\bibfnamefont {S.}~\bibnamefont
  {Bl\"ugel}}, \ and\ \bibinfo {author} {\bibfnamefont {C.}~\bibnamefont
  {Friedrich}},\ }\href {\doibase 10.1103/PhysRevB.100.045130} {\bibfield
  {journal} {\bibinfo  {journal} {Phys. Rev. B}\ }\textbf {\bibinfo {volume}
  {100}},\ \bibinfo {pages} {045130} (\bibinfo {year} {2019})}\BibitemShut
  {NoStop}%
\bibitem [{\citenamefont {Raquet}\ \emph {et~al.}(2002)\citenamefont {Raquet},
  \citenamefont {Viret}, \citenamefont {Sondergard}, \citenamefont {Cespedes},\
  and\ \citenamefont {Mamy}}]{Raquet2002}%
  \BibitemOpen
  \bibfield  {author} {\bibinfo {author} {\bibfnamefont {B.}~\bibnamefont
  {Raquet}}, \bibinfo {author} {\bibfnamefont {M.}~\bibnamefont {Viret}},
  \bibinfo {author} {\bibfnamefont {E.}~\bibnamefont {Sondergard}}, \bibinfo
  {author} {\bibfnamefont {O.}~\bibnamefont {Cespedes}}, \ and\ \bibinfo
  {author} {\bibfnamefont {R.}~\bibnamefont {Mamy}},\ }\href {\doibase
  10.1103/PhysRevB.66.024433} {\bibfield  {journal} {\bibinfo  {journal} {Phys.
  Rev. B}\ }\textbf {\bibinfo {volume} {66}},\ \bibinfo {pages} {024433}
  (\bibinfo {year} {2002})}\BibitemShut {NoStop}%
\bibitem [{\citenamefont {Giustino}(2017)}]{Giustino2017}%
  \BibitemOpen
  \bibfield  {author} {\bibinfo {author} {\bibfnamefont {F.}~\bibnamefont
  {Giustino}},\ }\href {\doibase 10.1103/RevModPhys.89.015003} {\bibfield
  {journal} {\bibinfo  {journal} {Rev. Mod. Phys.}\ }\textbf {\bibinfo {volume}
  {89}},\ \bibinfo {pages} {015003} (\bibinfo {year} {2017})}\BibitemShut
  {NoStop}%
\bibitem [{\citenamefont {Ponc\'e}\ \emph {et~al.}(2018)\citenamefont
  {Ponc\'e}, \citenamefont {Margine},\ and\ \citenamefont
  {Giustino}}]{Ponce2018}%
  \BibitemOpen
  \bibfield  {author} {\bibinfo {author} {\bibfnamefont {S.}~\bibnamefont
  {Ponc\'e}}, \bibinfo {author} {\bibfnamefont {E.~R.}\ \bibnamefont
  {Margine}}, \ and\ \bibinfo {author} {\bibfnamefont {F.}~\bibnamefont
  {Giustino}},\ }\href {\doibase 10.1103/PhysRevB.97.121201} {\bibfield
  {journal} {\bibinfo  {journal} {Phys. Rev. B}\ }\textbf {\bibinfo {volume}
  {97}},\ \bibinfo {pages} {121201} (\bibinfo {year} {2018})}\BibitemShut
  {NoStop}%
\bibitem [{\citenamefont {Sohier}\ \emph {et~al.}(2018)\citenamefont {Sohier},
  \citenamefont {Campi}, \citenamefont {Marzari},\ and\ \citenamefont
  {Gibertini}}]{Sohier2018}%
  \BibitemOpen
  \bibfield  {author} {\bibinfo {author} {\bibfnamefont {T.}~\bibnamefont
  {Sohier}}, \bibinfo {author} {\bibfnamefont {D.}~\bibnamefont {Campi}},
  \bibinfo {author} {\bibfnamefont {N.}~\bibnamefont {Marzari}}, \ and\
  \bibinfo {author} {\bibfnamefont {M.}~\bibnamefont {Gibertini}},\ }\href
  {\doibase 10.1103/PhysRevMaterials.2.114010} {\bibfield  {journal} {\bibinfo
  {journal} {Phys. Rev. Mater.}\ }\textbf {\bibinfo {volume} {2}},\ \bibinfo
  {pages} {114010} (\bibinfo {year} {2018})}\BibitemShut {NoStop}%
\bibitem [{\citenamefont {Lugovskoi}\ \emph {et~al.}(2019)\citenamefont
  {Lugovskoi}, \citenamefont {Katsnelson},\ and\ \citenamefont
  {Rudenko}}]{Lugovskoi2019}%
  \BibitemOpen
  \bibfield  {author} {\bibinfo {author} {\bibfnamefont {A.~V.}\ \bibnamefont
  {Lugovskoi}}, \bibinfo {author} {\bibfnamefont {M.~I.}\ \bibnamefont
  {Katsnelson}}, \ and\ \bibinfo {author} {\bibfnamefont {A.~N.}\ \bibnamefont
  {Rudenko}},\ }\href {\doibase 10.1103/PhysRevLett.123.176401} {\bibfield
  {journal} {\bibinfo  {journal} {Phys. Rev. Lett.}\ }\textbf {\bibinfo
  {volume} {123}},\ \bibinfo {pages} {176401} (\bibinfo {year}
  {2019})}\BibitemShut {NoStop}%
\bibitem [{\citenamefont {Poncé}\ \emph {et~al.}(2020)\citenamefont {Poncé},
  \citenamefont {Li}, \citenamefont {Reichardt},\ and\ \citenamefont
  {Giustino}}]{Ponce2020}%
  \BibitemOpen
  \bibfield  {author} {\bibinfo {author} {\bibfnamefont {S.}~\bibnamefont
  {Poncé}}, \bibinfo {author} {\bibfnamefont {W.}~\bibnamefont {Li}}, \bibinfo
  {author} {\bibfnamefont {S.}~\bibnamefont {Reichardt}}, \ and\ \bibinfo
  {author} {\bibfnamefont {F.}~\bibnamefont {Giustino}},\ }\href {\doibase
  10.1088/1361-6633/ab6a43} {\bibfield  {journal} {\bibinfo  {journal} {Reports
  on Progress in Physics}\ }\textbf {\bibinfo {volume} {83}},\ \bibinfo {pages}
  {036501} (\bibinfo {year} {2020})}\BibitemShut {NoStop}%
\bibitem [{\citenamefont {Rudenko}\ and\ \citenamefont
  {Yuan}(2020)}]{Rudenko2020}%
  \BibitemOpen
  \bibfield  {author} {\bibinfo {author} {\bibfnamefont {A.~N.}\ \bibnamefont
  {Rudenko}}\ and\ \bibinfo {author} {\bibfnamefont {S.}~\bibnamefont {Yuan}},\
  }\href {\doibase 10.1103/PhysRevB.101.115127} {\bibfield  {journal} {\bibinfo
   {journal} {Phys. Rev. B}\ }\textbf {\bibinfo {volume} {101}},\ \bibinfo
  {pages} {115127} (\bibinfo {year} {2020})}\BibitemShut {NoStop}%
\bibitem [{\citenamefont {Katsnelson}(2020)}]{Graphene_book}%
  \BibitemOpen
  \bibfield  {author} {\bibinfo {author} {\bibfnamefont {M.~I.}\ \bibnamefont
  {Katsnelson}},\ }\href@noop {} {\emph {\bibinfo {title} {The Physics of
  Graphene}}}\ (\bibinfo  {publisher} {Cambridge University Press},\ \bibinfo
  {year} {2020})\BibitemShut {NoStop}%
\bibitem [{\citenamefont {Rudenko}\ \emph {et~al.}(2016)\citenamefont
  {Rudenko}, \citenamefont {Brener},\ and\ \citenamefont
  {Katsnelson}}]{Rudenko2016}%
  \BibitemOpen
  \bibfield  {author} {\bibinfo {author} {\bibfnamefont {A.~N.}\ \bibnamefont
  {Rudenko}}, \bibinfo {author} {\bibfnamefont {S.}~\bibnamefont {Brener}}, \
  and\ \bibinfo {author} {\bibfnamefont {M.~I.}\ \bibnamefont {Katsnelson}},\
  }\href {\doibase 10.1103/PhysRevLett.116.246401} {\bibfield  {journal}
  {\bibinfo  {journal} {Phys. Rev. Lett.}\ }\textbf {\bibinfo {volume} {116}},\
  \bibinfo {pages} {246401} (\bibinfo {year} {2016})}\BibitemShut {NoStop}%
\bibitem [{\citenamefont {Fischetti}\ and\ \citenamefont
  {Vandenberghe}(2016)}]{PhysRevB.93.155413}%
  \BibitemOpen
  \bibfield  {author} {\bibinfo {author} {\bibfnamefont {M.~V.}\ \bibnamefont
  {Fischetti}}\ and\ \bibinfo {author} {\bibfnamefont {W.~G.}\ \bibnamefont
  {Vandenberghe}},\ }\href {\doibase 10.1103/PhysRevB.93.155413} {\bibfield
  {journal} {\bibinfo  {journal} {Phys. Rev. B}\ }\textbf {\bibinfo {volume}
  {93}},\ \bibinfo {pages} {155413} (\bibinfo {year} {2016})}\BibitemShut
  {NoStop}%
\bibitem [{\citenamefont {Rudenko}\ \emph {et~al.}(2019)\citenamefont
  {Rudenko}, \citenamefont {Lugovskoi}, \citenamefont {Mauri}, \citenamefont
  {Yu}, \citenamefont {Yuan},\ and\ \citenamefont {Katsnelson}}]{Rudenko2019}%
  \BibitemOpen
  \bibfield  {author} {\bibinfo {author} {\bibfnamefont {A.~N.}\ \bibnamefont
  {Rudenko}}, \bibinfo {author} {\bibfnamefont {A.~V.}\ \bibnamefont
  {Lugovskoi}}, \bibinfo {author} {\bibfnamefont {A.}~\bibnamefont {Mauri}},
  \bibinfo {author} {\bibfnamefont {G.}~\bibnamefont {Yu}}, \bibinfo {author}
  {\bibfnamefont {S.}~\bibnamefont {Yuan}}, \ and\ \bibinfo {author}
  {\bibfnamefont {M.~I.}\ \bibnamefont {Katsnelson}},\ }\href {\doibase
  10.1103/PhysRevB.100.075417} {\bibfield  {journal} {\bibinfo  {journal}
  {Phys. Rev. B}\ }\textbf {\bibinfo {volume} {100}},\ \bibinfo {pages}
  {075417} (\bibinfo {year} {2019})}\BibitemShut {NoStop}%
\bibitem [{\citenamefont {Nabok}\ \emph {et~al.}(2021)\citenamefont {Nabok},
  \citenamefont {Bl{\"u}gel},\ and\ \citenamefont {Friedrich}}]{Nabok2021}%
  \BibitemOpen
  \bibfield  {author} {\bibinfo {author} {\bibfnamefont {D.}~\bibnamefont
  {Nabok}}, \bibinfo {author} {\bibfnamefont {S.}~\bibnamefont {Bl{\"u}gel}}, \
  and\ \bibinfo {author} {\bibfnamefont {C.}~\bibnamefont {Friedrich}},\ }\href
  {\doibase 10.1038/s41524-021-00649-8} {\bibfield  {journal} {\bibinfo
  {journal} {npj Comput. Mater.}\ }\textbf {\bibinfo {volume} {7}},\ \bibinfo
  {pages} {178} (\bibinfo {year} {2021})}\BibitemShut {NoStop}%
\bibitem [{\citenamefont {Irkhin}\ and\ \citenamefont
  {Katsnelson}(1989)}]{Irkhin1989}%
  \BibitemOpen
  \bibfield  {author} {\bibinfo {author} {\bibfnamefont {V.~Y.}\ \bibnamefont
  {Irkhin}}\ and\ \bibinfo {author} {\bibfnamefont {M.~I.}\ \bibnamefont
  {Katsnelson}},\ }\href {\doibase 10.1007/BF01313569} {\bibfield  {journal}
  {\bibinfo  {journal} {Zeitschrift f\"{u}r Physik B Condensed Matter}\
  }\textbf {\bibinfo {volume} {75}},\ \bibinfo {pages} {67} (\bibinfo {year}
  {1989})}\BibitemShut {NoStop}%
\bibitem [{\citenamefont {Katsnelson}\ \emph {et~al.}(2008)\citenamefont
  {Katsnelson}, \citenamefont {Irkhin}, \citenamefont {Chioncel}, \citenamefont
  {Lichtenstein},\ and\ \citenamefont {de~Groot}}]{Katsnelson2008}%
  \BibitemOpen
  \bibfield  {author} {\bibinfo {author} {\bibfnamefont {M.~I.}\ \bibnamefont
  {Katsnelson}}, \bibinfo {author} {\bibfnamefont {V.~Y.}\ \bibnamefont
  {Irkhin}}, \bibinfo {author} {\bibfnamefont {L.}~\bibnamefont {Chioncel}},
  \bibinfo {author} {\bibfnamefont {A.~I.}\ \bibnamefont {Lichtenstein}}, \
  and\ \bibinfo {author} {\bibfnamefont {R.~A.}\ \bibnamefont {de~Groot}},\
  }\href {\doibase https://doi.org/10.1103/RevModPhys.80.315} {\bibfield
  {journal} {\bibinfo  {journal} {Reviews of Modern Physics}\ }\textbf
  {\bibinfo {volume} {80}},\ \bibinfo {pages} {315} (\bibinfo {year}
  {2008})}\BibitemShut {NoStop}%
\bibitem [{\citenamefont {Irkhin}\ and\ \citenamefont
  {Irkhin}(2007)}]{Irkhin2007_book}%
  \BibitemOpen
  \bibfield  {author} {\bibinfo {author} {\bibfnamefont {V.~Y.}\ \bibnamefont
  {Irkhin}}\ and\ \bibinfo {author} {\bibfnamefont {Y.~P.}\ \bibnamefont
  {Irkhin}},\ }\href {https://books.google.nl/books?id=R5Qr\_dxCbg4C} {\emph
  {\bibinfo {title} {Electronic Structure, Correlation Effects and Physical
  Properties of D- and F-metals and Their Compounds}}}\ (\bibinfo  {publisher}
  {Cambridge International Science Publishing, Limited},\ \bibinfo {year}
  {2007})\BibitemShut {NoStop}%
\bibitem [{\citenamefont {Feng}\ \emph {et~al.}(2022)\citenamefont {Feng},
  \citenamefont {Li}, \citenamefont {Shi}, \citenamefont {Xie}, \citenamefont
  {Li},\ and\ \citenamefont {Xu}}]{FGT_Transport_Feng2022}%
  \BibitemOpen
  \bibfield  {author} {\bibinfo {author} {\bibfnamefont {H.}~\bibnamefont
  {Feng}}, \bibinfo {author} {\bibfnamefont {Y.}~\bibnamefont {Li}}, \bibinfo
  {author} {\bibfnamefont {Y.}~\bibnamefont {Shi}}, \bibinfo {author}
  {\bibfnamefont {H.-Y.}\ \bibnamefont {Xie}}, \bibinfo {author} {\bibfnamefont
  {Y.}~\bibnamefont {Li}}, \ and\ \bibinfo {author} {\bibfnamefont
  {Y.}~\bibnamefont {Xu}},\ }\href {\doibase 10.1088/0256-307x/39/7/077501}
  {\bibfield  {journal} {\bibinfo  {journal} {Chinese Physics Letters}\
  }\textbf {\bibinfo {volume} {39}},\ \bibinfo {pages} {077501} (\bibinfo
  {year} {2022})}\BibitemShut {NoStop}%
\bibitem [{\citenamefont {Chen}\ \emph
  {et~al.}(2013{\natexlab{b}})\citenamefont {Chen}, \citenamefont {Yang},
  \citenamefont {Wang}, \citenamefont {Imai}, \citenamefont {Ohta},
  \citenamefont {Michioka}, \citenamefont {Yoshimura},\ and\ \citenamefont
  {Fang}}]{FGT_transport_Chen2013}%
  \BibitemOpen
  \bibfield  {author} {\bibinfo {author} {\bibfnamefont {B.}~\bibnamefont
  {Chen}}, \bibinfo {author} {\bibfnamefont {J.}~\bibnamefont {Yang}}, \bibinfo
  {author} {\bibfnamefont {H.}~\bibnamefont {Wang}}, \bibinfo {author}
  {\bibfnamefont {M.}~\bibnamefont {Imai}}, \bibinfo {author} {\bibfnamefont
  {H.}~\bibnamefont {Ohta}}, \bibinfo {author} {\bibfnamefont {C.}~\bibnamefont
  {Michioka}}, \bibinfo {author} {\bibfnamefont {K.}~\bibnamefont {Yoshimura}},
  \ and\ \bibinfo {author} {\bibfnamefont {M.}~\bibnamefont {Fang}},\ }\href
  {\doibase 10.7566/JPSJ.82.124711} {\bibfield  {journal} {\bibinfo  {journal}
  {Journal of the Physical Society of Japan}\ }\textbf {\bibinfo {volume}
  {82}},\ \bibinfo {pages} {124711} (\bibinfo {year}
  {2013}{\natexlab{b}})}\BibitemShut {NoStop}%
\bibitem [{\citenamefont {Liu}\ \emph {et~al.}(2018)\citenamefont {Liu},
  \citenamefont {Stavitski}, \citenamefont {Attenkofer},\ and\ \citenamefont
  {Petrovic}}]{FGT_transport_Liu2018}%
  \BibitemOpen
  \bibfield  {author} {\bibinfo {author} {\bibfnamefont {Y.}~\bibnamefont
  {Liu}}, \bibinfo {author} {\bibfnamefont {E.}~\bibnamefont {Stavitski}},
  \bibinfo {author} {\bibfnamefont {K.}~\bibnamefont {Attenkofer}}, \ and\
  \bibinfo {author} {\bibfnamefont {C.}~\bibnamefont {Petrovic}},\ }\href
  {\doibase 10.1103/PhysRevB.97.165415} {\bibfield  {journal} {\bibinfo
  {journal} {Phys. Rev. B}\ }\textbf {\bibinfo {volume} {97}},\ \bibinfo
  {pages} {165415} (\bibinfo {year} {2018})}\BibitemShut {NoStop}%
\bibitem [{\citenamefont {Deng}\ \emph
  {et~al.}(2018{\natexlab{b}})\citenamefont {Deng}, \citenamefont {Yu},
  \citenamefont {Song}, \citenamefont {Zhang}, \citenamefont {Wang},
  \citenamefont {Sun}, \citenamefont {Yi}, \citenamefont {Wu}, \citenamefont
  {Wu}, \citenamefont {Zhu}, \citenamefont {Wang}, \citenamefont {Chen},\ and\
  \citenamefont {Zhang}}]{Deng2018}%
  \BibitemOpen
  \bibfield  {author} {\bibinfo {author} {\bibfnamefont {Y.}~\bibnamefont
  {Deng}}, \bibinfo {author} {\bibfnamefont {Y.}~\bibnamefont {Yu}}, \bibinfo
  {author} {\bibfnamefont {Y.}~\bibnamefont {Song}}, \bibinfo {author}
  {\bibfnamefont {J.}~\bibnamefont {Zhang}}, \bibinfo {author} {\bibfnamefont
  {N.~Z.}\ \bibnamefont {Wang}}, \bibinfo {author} {\bibfnamefont
  {Z.}~\bibnamefont {Sun}}, \bibinfo {author} {\bibfnamefont {Y.}~\bibnamefont
  {Yi}}, \bibinfo {author} {\bibfnamefont {Y.~Z.}\ \bibnamefont {Wu}}, \bibinfo
  {author} {\bibfnamefont {S.}~\bibnamefont {Wu}}, \bibinfo {author}
  {\bibfnamefont {J.}~\bibnamefont {Zhu}}, \bibinfo {author} {\bibfnamefont
  {J.}~\bibnamefont {Wang}}, \bibinfo {author} {\bibfnamefont {X.~H.}\
  \bibnamefont {Chen}}, \ and\ \bibinfo {author} {\bibfnamefont
  {Y.}~\bibnamefont {Zhang}},\ }\href {\doibase 10.1038/s41586-018-0626-9}
  {\bibfield  {journal} {\bibinfo  {journal} {Nature}\ }\textbf {\bibinfo
  {volume} {563}},\ \bibinfo {pages} {94} (\bibinfo {year}
  {2018}{\natexlab{b}})}\BibitemShut {NoStop}%
\bibitem [{\citenamefont {Giannozzi}\ \emph {et~al.}(2020)\citenamefont
  {Giannozzi}, \citenamefont {Baseggio}, \citenamefont {Bonfà}, \citenamefont
  {Brunato}, \citenamefont {Car}, \citenamefont {Carnimeo}, \citenamefont
  {Cavazzoni}, \citenamefont {de~Gironcoli}, \citenamefont {Delugas},
  \citenamefont {Ferrari~Ruffino}, \citenamefont {Ferretti}, \citenamefont
  {Marzari}, \citenamefont {Timrov}, \citenamefont {Urru},\ and\ \citenamefont
  {Baroni}}]{QE1}%
  \BibitemOpen
  \bibfield  {author} {\bibinfo {author} {\bibfnamefont {P.}~\bibnamefont
  {Giannozzi}}, \bibinfo {author} {\bibfnamefont {O.}~\bibnamefont {Baseggio}},
  \bibinfo {author} {\bibfnamefont {P.}~\bibnamefont {Bonfà}}, \bibinfo
  {author} {\bibfnamefont {D.}~\bibnamefont {Brunato}}, \bibinfo {author}
  {\bibfnamefont {R.}~\bibnamefont {Car}}, \bibinfo {author} {\bibfnamefont
  {I.}~\bibnamefont {Carnimeo}}, \bibinfo {author} {\bibfnamefont
  {C.}~\bibnamefont {Cavazzoni}}, \bibinfo {author} {\bibfnamefont
  {S.}~\bibnamefont {de~Gironcoli}}, \bibinfo {author} {\bibfnamefont
  {P.}~\bibnamefont {Delugas}}, \bibinfo {author} {\bibfnamefont
  {F.}~\bibnamefont {Ferrari~Ruffino}}, \bibinfo {author} {\bibfnamefont
  {A.}~\bibnamefont {Ferretti}}, \bibinfo {author} {\bibfnamefont
  {N.}~\bibnamefont {Marzari}}, \bibinfo {author} {\bibfnamefont
  {I.}~\bibnamefont {Timrov}}, \bibinfo {author} {\bibfnamefont
  {A.}~\bibnamefont {Urru}}, \ and\ \bibinfo {author} {\bibfnamefont
  {S.}~\bibnamefont {Baroni}},\ }\href {\doibase 10.1063/5.0005082} {\bibfield
  {journal} {\bibinfo  {journal} {The Journal of Chemical Physics}\ }\textbf
  {\bibinfo {volume} {152}},\ \bibinfo {pages} {154105} (\bibinfo {year}
  {2020})}\BibitemShut {NoStop}%
\bibitem [{\citenamefont {Giannozzi}\ \emph {et~al.}(2017)\citenamefont
  {Giannozzi}, \citenamefont {Andreussi}, \citenamefont {Brumme}, \citenamefont
  {Bunau}, \citenamefont {Nardelli}, \citenamefont {Calandra}, \citenamefont
  {Car}, \citenamefont {Cavazzoni}, \citenamefont {Ceresoli}, \citenamefont
  {Cococcioni}, \citenamefont {Colonna}, \citenamefont {Carnimeo},
  \citenamefont {Corso}, \citenamefont {de~Gironcoli}, \citenamefont {Delugas},
  \citenamefont {DiStasio}, \citenamefont {Ferretti}, \citenamefont {Floris},
  \citenamefont {Fratesi}, \citenamefont {Fugallo}, \citenamefont {Gebauer},
  \citenamefont {Gerstmann}, \citenamefont {Giustino}, \citenamefont {Gorni},
  \citenamefont {Jia}, \citenamefont {Kawamura}, \citenamefont {Ko},
  \citenamefont {Kokalj}, \citenamefont {Kü{\c{c}}ükbenli}, \citenamefont
  {Lazzeri}, \citenamefont {Marsili}, \citenamefont {Marzari}, \citenamefont
  {Mauri}, \citenamefont {Nguyen}, \citenamefont {Nguyen}, \citenamefont {de-la
  Roza}, \citenamefont {Paulatto}, \citenamefont {Ponc{\'{e}}}, \citenamefont
  {Rocca}, \citenamefont {Sabatini}, \citenamefont {Santra}, \citenamefont
  {Schlipf}, \citenamefont {Seitsonen}, \citenamefont {Smogunov}, \citenamefont
  {Timrov}, \citenamefont {Thonhauser}, \citenamefont {Umari}, \citenamefont
  {Vast}, \citenamefont {Wu},\ and\ \citenamefont {Baroni}}]{QE2}%
  \BibitemOpen
  \bibfield  {author} {\bibinfo {author} {\bibfnamefont {P.}~\bibnamefont
  {Giannozzi}}, \bibinfo {author} {\bibfnamefont {O.}~\bibnamefont
  {Andreussi}}, \bibinfo {author} {\bibfnamefont {T.}~\bibnamefont {Brumme}},
  \bibinfo {author} {\bibfnamefont {O.}~\bibnamefont {Bunau}}, \bibinfo
  {author} {\bibfnamefont {M.~B.}\ \bibnamefont {Nardelli}}, \bibinfo {author}
  {\bibfnamefont {M.}~\bibnamefont {Calandra}}, \bibinfo {author}
  {\bibfnamefont {R.}~\bibnamefont {Car}}, \bibinfo {author} {\bibfnamefont
  {C.}~\bibnamefont {Cavazzoni}}, \bibinfo {author} {\bibfnamefont
  {D.}~\bibnamefont {Ceresoli}}, \bibinfo {author} {\bibfnamefont
  {M.}~\bibnamefont {Cococcioni}}, \bibinfo {author} {\bibfnamefont
  {N.}~\bibnamefont {Colonna}}, \bibinfo {author} {\bibfnamefont
  {I.}~\bibnamefont {Carnimeo}}, \bibinfo {author} {\bibfnamefont {A.~D.}\
  \bibnamefont {Corso}}, \bibinfo {author} {\bibfnamefont {S.}~\bibnamefont
  {de~Gironcoli}}, \bibinfo {author} {\bibfnamefont {P.}~\bibnamefont
  {Delugas}}, \bibinfo {author} {\bibfnamefont {R.~A.}\ \bibnamefont
  {DiStasio}}, \bibinfo {author} {\bibfnamefont {A.}~\bibnamefont {Ferretti}},
  \bibinfo {author} {\bibfnamefont {A.}~\bibnamefont {Floris}}, \bibinfo
  {author} {\bibfnamefont {G.}~\bibnamefont {Fratesi}}, \bibinfo {author}
  {\bibfnamefont {G.}~\bibnamefont {Fugallo}}, \bibinfo {author} {\bibfnamefont
  {R.}~\bibnamefont {Gebauer}}, \bibinfo {author} {\bibfnamefont
  {U.}~\bibnamefont {Gerstmann}}, \bibinfo {author} {\bibfnamefont
  {F.}~\bibnamefont {Giustino}}, \bibinfo {author} {\bibfnamefont
  {T.}~\bibnamefont {Gorni}}, \bibinfo {author} {\bibfnamefont
  {J.}~\bibnamefont {Jia}}, \bibinfo {author} {\bibfnamefont {M.}~\bibnamefont
  {Kawamura}}, \bibinfo {author} {\bibfnamefont {H.-Y.}\ \bibnamefont {Ko}},
  \bibinfo {author} {\bibfnamefont {A.}~\bibnamefont {Kokalj}}, \bibinfo
  {author} {\bibfnamefont {E.}~\bibnamefont {Kü{\c{c}}ükbenli}}, \bibinfo
  {author} {\bibfnamefont {M.}~\bibnamefont {Lazzeri}}, \bibinfo {author}
  {\bibfnamefont {M.}~\bibnamefont {Marsili}}, \bibinfo {author} {\bibfnamefont
  {N.}~\bibnamefont {Marzari}}, \bibinfo {author} {\bibfnamefont
  {F.}~\bibnamefont {Mauri}}, \bibinfo {author} {\bibfnamefont {N.~L.}\
  \bibnamefont {Nguyen}}, \bibinfo {author} {\bibfnamefont {H.-V.}\
  \bibnamefont {Nguyen}}, \bibinfo {author} {\bibfnamefont {A.~O.}\
  \bibnamefont {de-la Roza}}, \bibinfo {author} {\bibfnamefont
  {L.}~\bibnamefont {Paulatto}}, \bibinfo {author} {\bibfnamefont
  {S.}~\bibnamefont {Ponc{\'{e}}}}, \bibinfo {author} {\bibfnamefont
  {D.}~\bibnamefont {Rocca}}, \bibinfo {author} {\bibfnamefont
  {R.}~\bibnamefont {Sabatini}}, \bibinfo {author} {\bibfnamefont
  {B.}~\bibnamefont {Santra}}, \bibinfo {author} {\bibfnamefont
  {M.}~\bibnamefont {Schlipf}}, \bibinfo {author} {\bibfnamefont {A.~P.}\
  \bibnamefont {Seitsonen}}, \bibinfo {author} {\bibfnamefont {A.}~\bibnamefont
  {Smogunov}}, \bibinfo {author} {\bibfnamefont {I.}~\bibnamefont {Timrov}},
  \bibinfo {author} {\bibfnamefont {T.}~\bibnamefont {Thonhauser}}, \bibinfo
  {author} {\bibfnamefont {P.}~\bibnamefont {Umari}}, \bibinfo {author}
  {\bibfnamefont {N.}~\bibnamefont {Vast}}, \bibinfo {author} {\bibfnamefont
  {X.}~\bibnamefont {Wu}}, \ and\ \bibinfo {author} {\bibfnamefont
  {S.}~\bibnamefont {Baroni}},\ }\href {\doibase 10.1088/1361-648x/aa8f79}
  {\bibfield  {journal} {\bibinfo  {journal} {Journal of Physics: Condensed
  Matter}\ }\textbf {\bibinfo {volume} {29}},\ \bibinfo {pages} {465901}
  (\bibinfo {year} {2017})}\BibitemShut {NoStop}%
\bibitem [{\citenamefont {Perdew}\ \emph {et~al.}(1996)\citenamefont {Perdew},
  \citenamefont {Burke},\ and\ \citenamefont {Ernzerhof}}]{PBE}%
  \BibitemOpen
  \bibfield  {author} {\bibinfo {author} {\bibfnamefont {J.~P.}\ \bibnamefont
  {Perdew}}, \bibinfo {author} {\bibfnamefont {K.}~\bibnamefont {Burke}}, \
  and\ \bibinfo {author} {\bibfnamefont {M.}~\bibnamefont {Ernzerhof}},\ }\href
  {https://link.aps.org/doi/10.1103/PhysRevLett.77.3865} {\bibfield  {journal}
  {\bibinfo  {journal} {Phys. Rev. Lett.}\ }\textbf {\bibinfo {volume} {77}},\
  \bibinfo {pages} {3865} (\bibinfo {year} {1996})}\BibitemShut {NoStop}%
\bibitem [{\citenamefont {Deiseroth}\ \emph
  {et~al.}(2006{\natexlab{b}})\citenamefont {Deiseroth}, \citenamefont
  {Aleksandrov}, \citenamefont {Reiner}, \citenamefont {Kienle},\ and\
  \citenamefont {Kremer}}]{Deiseroth2006}%
  \BibitemOpen
  \bibfield  {author} {\bibinfo {author} {\bibfnamefont {H.-J.}\ \bibnamefont
  {Deiseroth}}, \bibinfo {author} {\bibfnamefont {K.}~\bibnamefont
  {Aleksandrov}}, \bibinfo {author} {\bibfnamefont {C.}~\bibnamefont {Reiner}},
  \bibinfo {author} {\bibfnamefont {L.}~\bibnamefont {Kienle}}, \ and\ \bibinfo
  {author} {\bibfnamefont {R.~K.}\ \bibnamefont {Kremer}},\ }\href {\doibase
  https://doi.org/10.1002/ejic.200501020} {\bibfield  {journal} {\bibinfo
  {journal} {European Journal of Inorganic Chemistry}\ }\textbf {\bibinfo
  {volume} {2006}},\ \bibinfo {pages} {1561} (\bibinfo {year}
  {2006}{\natexlab{b}})}\BibitemShut {NoStop}%
\bibitem [{\citenamefont {Marzari}\ and\ \citenamefont
  {Vanderbilt}(1997)}]{marzari1997}%
  \BibitemOpen
  \bibfield  {author} {\bibinfo {author} {\bibfnamefont {N.}~\bibnamefont
  {Marzari}}\ and\ \bibinfo {author} {\bibfnamefont {D.}~\bibnamefont
  {Vanderbilt}},\ }\href {\doibase 10.1103/PhysRevB.56.12847} {\bibfield
  {journal} {\bibinfo  {journal} {Phys. Rev. B}\ }\textbf {\bibinfo {volume}
  {56}},\ \bibinfo {pages} {12847} (\bibinfo {year} {1997})}\BibitemShut
  {NoStop}%
\bibitem [{\citenamefont {Pizzi}\ \emph {et~al.}(2020)\citenamefont {Pizzi},
  \citenamefont {Vitale}, \citenamefont {Arita}, \citenamefont {Bl\"ugel},
  \citenamefont {Freimuth}, \citenamefont {G{\'{e}}ranton}, \citenamefont
  {Gibertini}, \citenamefont {Gresch}, \citenamefont {Johnson}, \citenamefont
  {Koretsune}, \citenamefont {Iba{\~{n}}ez-Azpiroz}, \citenamefont {Lee},
  \citenamefont {Lihm}, \citenamefont {Marchand}, \citenamefont {Marrazzo},
  \citenamefont {Mokrousov}, \citenamefont {Mustafa}, \citenamefont {Nohara},
  \citenamefont {Nomura}, \citenamefont {Paulatto}, \citenamefont
  {Ponc{\'{e}}}, \citenamefont {Ponweiser}, \citenamefont {Qiao}, \citenamefont
  {Th\"ole}, \citenamefont {Tsirkin}, \citenamefont {Wierzbowska},
  \citenamefont {Marzari}, \citenamefont {Vanderbilt}, \citenamefont {Souza},
  \citenamefont {Mostofi},\ and\ \citenamefont {Yates}}]{pizzi2020}%
  \BibitemOpen
  \bibfield  {author} {\bibinfo {author} {\bibfnamefont {G.}~\bibnamefont
  {Pizzi}}, \bibinfo {author} {\bibfnamefont {V.}~\bibnamefont {Vitale}},
  \bibinfo {author} {\bibfnamefont {R.}~\bibnamefont {Arita}}, \bibinfo
  {author} {\bibfnamefont {S.}~\bibnamefont {Bl\"ugel}}, \bibinfo {author}
  {\bibfnamefont {F.}~\bibnamefont {Freimuth}}, \bibinfo {author}
  {\bibfnamefont {G.}~\bibnamefont {G{\'{e}}ranton}}, \bibinfo {author}
  {\bibfnamefont {M.}~\bibnamefont {Gibertini}}, \bibinfo {author}
  {\bibfnamefont {D.}~\bibnamefont {Gresch}}, \bibinfo {author} {\bibfnamefont
  {C.}~\bibnamefont {Johnson}}, \bibinfo {author} {\bibfnamefont
  {T.}~\bibnamefont {Koretsune}}, \bibinfo {author} {\bibfnamefont
  {J.}~\bibnamefont {Iba{\~{n}}ez-Azpiroz}}, \bibinfo {author} {\bibfnamefont
  {H.}~\bibnamefont {Lee}}, \bibinfo {author} {\bibfnamefont {J.-M.}\
  \bibnamefont {Lihm}}, \bibinfo {author} {\bibfnamefont {D.}~\bibnamefont
  {Marchand}}, \bibinfo {author} {\bibfnamefont {A.}~\bibnamefont {Marrazzo}},
  \bibinfo {author} {\bibfnamefont {Y.}~\bibnamefont {Mokrousov}}, \bibinfo
  {author} {\bibfnamefont {J.~I.}\ \bibnamefont {Mustafa}}, \bibinfo {author}
  {\bibfnamefont {Y.}~\bibnamefont {Nohara}}, \bibinfo {author} {\bibfnamefont
  {Y.}~\bibnamefont {Nomura}}, \bibinfo {author} {\bibfnamefont
  {L.}~\bibnamefont {Paulatto}}, \bibinfo {author} {\bibfnamefont
  {S.}~\bibnamefont {Ponc{\'{e}}}}, \bibinfo {author} {\bibfnamefont
  {T.}~\bibnamefont {Ponweiser}}, \bibinfo {author} {\bibfnamefont
  {J.}~\bibnamefont {Qiao}}, \bibinfo {author} {\bibfnamefont {F.}~\bibnamefont
  {Th\"ole}}, \bibinfo {author} {\bibfnamefont {S.~S.}\ \bibnamefont
  {Tsirkin}}, \bibinfo {author} {\bibfnamefont {M.}~\bibnamefont
  {Wierzbowska}}, \bibinfo {author} {\bibfnamefont {N.}~\bibnamefont
  {Marzari}}, \bibinfo {author} {\bibfnamefont {D.}~\bibnamefont {Vanderbilt}},
  \bibinfo {author} {\bibfnamefont {I.}~\bibnamefont {Souza}}, \bibinfo
  {author} {\bibfnamefont {A.~A.}\ \bibnamefont {Mostofi}}, \ and\ \bibinfo
  {author} {\bibfnamefont {J.~R.}\ \bibnamefont {Yates}},\ }\href {\doibase
  10.1088/1361-648x/ab51ff} {\bibfield  {journal} {\bibinfo  {journal} {Journal
  of Physics: Condensed Matter}\ }\textbf {\bibinfo {volume} {32}},\ \bibinfo
  {pages} {165902} (\bibinfo {year} {2020})}\BibitemShut {NoStop}%
\bibitem [{\citenamefont {Baroni}\ \emph {et~al.}(2001)\citenamefont {Baroni},
  \citenamefont {de~Gironcoli}, \citenamefont {Dal~Corso},\ and\ \citenamefont
  {Giannozzi}}]{Baroni2001}%
  \BibitemOpen
  \bibfield  {author} {\bibinfo {author} {\bibfnamefont {S.}~\bibnamefont
  {Baroni}}, \bibinfo {author} {\bibfnamefont {S.}~\bibnamefont
  {de~Gironcoli}}, \bibinfo {author} {\bibfnamefont {A.}~\bibnamefont
  {Dal~Corso}}, \ and\ \bibinfo {author} {\bibfnamefont {P.}~\bibnamefont
  {Giannozzi}},\ }\href {\doibase 10.1103/RevModPhys.73.515} {\bibfield
  {journal} {\bibinfo  {journal} {Rev. Mod. Phys.}\ }\textbf {\bibinfo {volume}
  {73}},\ \bibinfo {pages} {515} (\bibinfo {year} {2001})}\BibitemShut
  {NoStop}%
\bibitem [{\citenamefont {Ziman}(1964)}]{Ziman1964_book}%
  \BibitemOpen
  \bibfield  {author} {\bibinfo {author} {\bibfnamefont {J.~M.}\ \bibnamefont
  {Ziman}},\ }\href@noop {} {\emph {\bibinfo {title} {{Principles of the Theory
  of Solids}}}}\ (\bibinfo  {publisher} {Cambridge University Press},\ \bibinfo
  {address} {Cambridge, England},\ \bibinfo {year} {1964})\BibitemShut
  {NoStop}%
\bibitem [{\citenamefont {Ziman}(2001)}]{Ziman2001_book}%
  \BibitemOpen
  \bibfield  {author} {\bibinfo {author} {\bibfnamefont {J.}~\bibnamefont
  {Ziman}},\ }\href@noop {} {\emph {\bibinfo {title} {{Electrons and Phonons:
  The Theory of Transport Phenomena in Solids}}}}\ (\bibinfo  {publisher}
  {Oxford University Press},\ \bibinfo {address} {Oxford, England},\ \bibinfo
  {year} {2001})\BibitemShut {NoStop}%
\bibitem [{\citenamefont {Ponc\'e}\ \emph {et~al.}(2016)\citenamefont
  {Ponc\'e}, \citenamefont {Margine}, \citenamefont {Verdi},\ and\
  \citenamefont {Giustino}}]{Ponce2016}%
  \BibitemOpen
  \bibfield  {author} {\bibinfo {author} {\bibfnamefont {S.}~\bibnamefont
  {Ponc\'e}}, \bibinfo {author} {\bibfnamefont {E.}~\bibnamefont {Margine}},
  \bibinfo {author} {\bibfnamefont {C.}~\bibnamefont {Verdi}}, \ and\ \bibinfo
  {author} {\bibfnamefont {F.}~\bibnamefont {Giustino}},\ }\href {\doibase
  https://doi.org/10.1016/j.cpc.2016.07.028} {\bibfield  {journal} {\bibinfo
  {journal} {Computer Physics Communications}\ }\textbf {\bibinfo {volume}
  {209}},\ \bibinfo {pages} {116} (\bibinfo {year} {2016})}\BibitemShut
  {NoStop}%
\bibitem [{\citenamefont {Liechtenstein}\ \emph {et~al.}(1987)\citenamefont
  {Liechtenstein}, \citenamefont {Katsnelson}, \citenamefont {Antropov},\ and\
  \citenamefont {Gubanov}}]{liechtenstein1987}%
  \BibitemOpen
  \bibfield  {author} {\bibinfo {author} {\bibfnamefont {A.~I.}\ \bibnamefont
  {Liechtenstein}}, \bibinfo {author} {\bibfnamefont {M.~I.}\ \bibnamefont
  {Katsnelson}}, \bibinfo {author} {\bibfnamefont {V.~P.}\ \bibnamefont
  {Antropov}}, \ and\ \bibinfo {author} {\bibfnamefont {V.~A.}\ \bibnamefont
  {Gubanov}},\ }\href {\doibase 10.1016/0304-8853(87)90721-9} {\bibfield
  {journal} {\bibinfo  {journal} {Journal of Magnetism and Magnetic Materials}\
  }\textbf {\bibinfo {volume} {67}},\ \bibinfo {pages} {65 } (\bibinfo {year}
  {1987})}\BibitemShut {NoStop}%
\bibitem [{\citenamefont {Kashin}\ \emph {et~al.}(2020)\citenamefont {Kashin},
  \citenamefont {Mazurenko}, \citenamefont {Katsnelson},\ and\ \citenamefont
  {Rudenko}}]{Kashin2020}%
  \BibitemOpen
  \bibfield  {author} {\bibinfo {author} {\bibfnamefont {I.~V.}\ \bibnamefont
  {Kashin}}, \bibinfo {author} {\bibfnamefont {V.~V.}\ \bibnamefont
  {Mazurenko}}, \bibinfo {author} {\bibfnamefont {M.~I.}\ \bibnamefont
  {Katsnelson}}, \ and\ \bibinfo {author} {\bibfnamefont {A.~N.}\ \bibnamefont
  {Rudenko}},\ }\href {\doibase 10.1088/2053-1583/ab72d8} {\bibfield  {journal}
  {\bibinfo  {journal} {2D Materials}\ }\textbf {\bibinfo {volume} {7}},\
  \bibinfo {pages} {025036} (\bibinfo {year} {2020})}\BibitemShut {NoStop}%
\bibitem [{\citenamefont {Rusz}\ \emph {et~al.}(2005)\citenamefont {Rusz},
  \citenamefont {Turek},\ and\ \citenamefont {Divi\ifmmode~\check{s}\else
  \v{s}\fi{}}}]{Rusz2005}%
  \BibitemOpen
  \bibfield  {author} {\bibinfo {author} {\bibfnamefont {J.}~\bibnamefont
  {Rusz}}, \bibinfo {author} {\bibfnamefont {I.}~\bibnamefont {Turek}}, \ and\
  \bibinfo {author} {\bibfnamefont {M.}~\bibnamefont
  {Divi\ifmmode~\check{s}\else \v{s}\fi{}}},\ }\href {\doibase
  10.1103/PhysRevB.71.174408} {\bibfield  {journal} {\bibinfo  {journal} {Phys.
  Rev. B}\ }\textbf {\bibinfo {volume} {71}},\ \bibinfo {pages} {174408}
  (\bibinfo {year} {2005})}\BibitemShut {NoStop}%
\bibitem [{\citenamefont {Tyablikov}(1983)}]{Tyablikov_book}%
  \BibitemOpen
  \bibfield  {author} {\bibinfo {author} {\bibfnamefont {S.~V.}\ \bibnamefont
  {Tyablikov}},\ }\href@noop {} {\emph {\bibinfo {title} {Methods in Quantum
  Theory of Magnetism}}}\ (\bibinfo  {publisher} {Springer, New York},\
  \bibinfo {year} {1983})\BibitemShut {NoStop}%
\bibitem [{\citenamefont {Nolting}\ and\ \citenamefont
  {Ramakanth}(2009)}]{Nolting_book}%
  \BibitemOpen
  \bibfield  {author} {\bibinfo {author} {\bibfnamefont {W.}~\bibnamefont
  {Nolting}}\ and\ \bibinfo {author} {\bibfnamefont {A.}~\bibnamefont
  {Ramakanth}},\ }\href@noop {} {\emph {\bibinfo {title} {Quantum Theory of
  Magnetism}}}\ (\bibinfo  {publisher} {Springer-Verlag, Berlin, Heidelberg},\
  \bibinfo {year} {2009})\BibitemShut {NoStop}%
\bibitem [{\citenamefont {Zhang}\ \emph {et~al.}(2018)\citenamefont {Zhang},
  \citenamefont {Lu}, \citenamefont {Zhu}, \citenamefont {Tan}, \citenamefont
  {Feng}, \citenamefont {Liu}, \citenamefont {Zhang}, \citenamefont {Chen},
  \citenamefont {Liu}, \citenamefont {Luo}, \citenamefont {Xie}, \citenamefont
  {Luo}, \citenamefont {Zhang},\ and\ \citenamefont
  {Lai}}]{FGT_transport_Zhang2018}%
  \BibitemOpen
  \bibfield  {author} {\bibinfo {author} {\bibfnamefont {Y.}~\bibnamefont
  {Zhang}}, \bibinfo {author} {\bibfnamefont {H.}~\bibnamefont {Lu}}, \bibinfo
  {author} {\bibfnamefont {X.}~\bibnamefont {Zhu}}, \bibinfo {author}
  {\bibfnamefont {S.}~\bibnamefont {Tan}}, \bibinfo {author} {\bibfnamefont
  {W.}~\bibnamefont {Feng}}, \bibinfo {author} {\bibfnamefont {Q.}~\bibnamefont
  {Liu}}, \bibinfo {author} {\bibfnamefont {W.}~\bibnamefont {Zhang}}, \bibinfo
  {author} {\bibfnamefont {Q.}~\bibnamefont {Chen}}, \bibinfo {author}
  {\bibfnamefont {Y.}~\bibnamefont {Liu}}, \bibinfo {author} {\bibfnamefont
  {X.}~\bibnamefont {Luo}}, \bibinfo {author} {\bibfnamefont {D.}~\bibnamefont
  {Xie}}, \bibinfo {author} {\bibfnamefont {L.}~\bibnamefont {Luo}}, \bibinfo
  {author} {\bibfnamefont {Z.}~\bibnamefont {Zhang}}, \ and\ \bibinfo {author}
  {\bibfnamefont {X.}~\bibnamefont {Lai}},\ }\href {\doibase
  10.1126/sciadv.aao6791} {\bibfield  {journal} {\bibinfo  {journal} {Science
  Advances}\ }\textbf {\bibinfo {volume} {4}},\ \bibinfo {pages} {eaao6791}
  (\bibinfo {year} {2018})}\BibitemShut {NoStop}%
\bibitem [{\citenamefont {Zhuang}\ \emph {et~al.}(2016)\citenamefont {Zhuang},
  \citenamefont {Kent},\ and\ \citenamefont {Hennig}}]{Zhuang2016}%
  \BibitemOpen
  \bibfield  {author} {\bibinfo {author} {\bibfnamefont {H.~L.}\ \bibnamefont
  {Zhuang}}, \bibinfo {author} {\bibfnamefont {P.~R.~C.}\ \bibnamefont {Kent}},
  \ and\ \bibinfo {author} {\bibfnamefont {R.~G.}\ \bibnamefont {Hennig}},\
  }\href {\doibase 10.1103/PhysRevB.93.134407} {\bibfield  {journal} {\bibinfo
  {journal} {Phys. Rev. B}\ }\textbf {\bibinfo {volume} {93}},\ \bibinfo
  {pages} {134407} (\bibinfo {year} {2016})}\BibitemShut {NoStop}%
\bibitem [{\citenamefont {Gyorffy}\ \emph {et~al.}(1985)\citenamefont
  {Gyorffy}, \citenamefont {Pindor}, \citenamefont {Staunton}, \citenamefont
  {Stocks},\ and\ \citenamefont {Winter}}]{Gyorffy_1985}%
  \BibitemOpen
  \bibfield  {author} {\bibinfo {author} {\bibfnamefont {B.~L.}\ \bibnamefont
  {Gyorffy}}, \bibinfo {author} {\bibfnamefont {A.~J.}\ \bibnamefont {Pindor}},
  \bibinfo {author} {\bibfnamefont {J.}~\bibnamefont {Staunton}}, \bibinfo
  {author} {\bibfnamefont {G.~M.}\ \bibnamefont {Stocks}}, \ and\ \bibinfo
  {author} {\bibfnamefont {H.}~\bibnamefont {Winter}},\ }\href {\doibase
  10.1088/0305-4608/15/6/018} {\bibfield  {journal} {\bibinfo  {journal}
  {Journal of Physics F: Metal Physics}\ }\textbf {\bibinfo {volume} {15}},\
  \bibinfo {pages} {1337} (\bibinfo {year} {1985})}\BibitemShut {NoStop}%
\bibitem [{\citenamefont {Staunton}\ \emph {et~al.}(1985)\citenamefont
  {Staunton}, \citenamefont {Gyorffy}, \citenamefont {Pindor}, \citenamefont
  {Stocks},\ and\ \citenamefont {Winter}}]{Staunton_1985}%
  \BibitemOpen
  \bibfield  {author} {\bibinfo {author} {\bibfnamefont {J.}~\bibnamefont
  {Staunton}}, \bibinfo {author} {\bibfnamefont {B.~L.}\ \bibnamefont
  {Gyorffy}}, \bibinfo {author} {\bibfnamefont {A.~J.}\ \bibnamefont {Pindor}},
  \bibinfo {author} {\bibfnamefont {G.~M.}\ \bibnamefont {Stocks}}, \ and\
  \bibinfo {author} {\bibfnamefont {H.}~\bibnamefont {Winter}},\ }\href
  {\doibase 10.1088/0305-4608/15/6/019} {\bibfield  {journal} {\bibinfo
  {journal} {Journal of Physics F: Metal Physics}\ }\textbf {\bibinfo {volume}
  {15}},\ \bibinfo {pages} {1387} (\bibinfo {year} {1985})}\BibitemShut
  {NoStop}%
\bibitem [{\citenamefont {Pindor}\ \emph {et~al.}(1983)\citenamefont {Pindor},
  \citenamefont {Staunton}, \citenamefont {Stocks},\ and\ \citenamefont
  {Winter}}]{Pindor_1983}%
  \BibitemOpen
  \bibfield  {author} {\bibinfo {author} {\bibfnamefont {A.~J.}\ \bibnamefont
  {Pindor}}, \bibinfo {author} {\bibfnamefont {J.}~\bibnamefont {Staunton}},
  \bibinfo {author} {\bibfnamefont {G.~M.}\ \bibnamefont {Stocks}}, \ and\
  \bibinfo {author} {\bibfnamefont {H.}~\bibnamefont {Winter}},\ }\href
  {\doibase 10.1088/0305-4608/13/5/012} {\bibfield  {journal} {\bibinfo
  {journal} {J. Phys. F: Met. Phys.}\ }\textbf {\bibinfo {volume} {13}},\
  \bibinfo {pages} {979} (\bibinfo {year} {1983})}\BibitemShut {NoStop}%
\bibitem [{\citenamefont {Staunton}\ \emph {et~al.}(1986)\citenamefont
  {Staunton}, \citenamefont {Gyorffy}, \citenamefont {Stocks},\ and\
  \citenamefont {Wadsworth}}]{Staunton_1986}%
  \BibitemOpen
  \bibfield  {author} {\bibinfo {author} {\bibfnamefont {J.}~\bibnamefont
  {Staunton}}, \bibinfo {author} {\bibfnamefont {B.~L.}\ \bibnamefont
  {Gyorffy}}, \bibinfo {author} {\bibfnamefont {G.~M.}\ \bibnamefont {Stocks}},
  \ and\ \bibinfo {author} {\bibfnamefont {J.}~\bibnamefont {Wadsworth}},\
  }\href {\doibase 10.1088/0305-4608/16/11/016} {\bibfield  {journal} {\bibinfo
   {journal} {J. Phys. F: Met. Phys.}\ }\textbf {\bibinfo {volume} {16}},\
  \bibinfo {pages} {1761} (\bibinfo {year} {1986})}\BibitemShut {NoStop}%
\bibitem [{\citenamefont {Niklasson}\ \emph {et~al.}(2003)\citenamefont
  {Niklasson}, \citenamefont {Wills}, \citenamefont {Katsnelson}, \citenamefont
  {Abrikosov}, \citenamefont {Eriksson},\ and\ \citenamefont
  {Johansson}}]{PhysRevB.67.235105}%
  \BibitemOpen
  \bibfield  {author} {\bibinfo {author} {\bibfnamefont {A.~M.~N.}\
  \bibnamefont {Niklasson}}, \bibinfo {author} {\bibfnamefont {J.~M.}\
  \bibnamefont {Wills}}, \bibinfo {author} {\bibfnamefont {M.~I.}\ \bibnamefont
  {Katsnelson}}, \bibinfo {author} {\bibfnamefont {I.~A.}\ \bibnamefont
  {Abrikosov}}, \bibinfo {author} {\bibfnamefont {O.}~\bibnamefont {Eriksson}},
  \ and\ \bibinfo {author} {\bibfnamefont {B.}~\bibnamefont {Johansson}},\
  }\href {\doibase 10.1103/PhysRevB.67.235105} {\bibfield  {journal} {\bibinfo
  {journal} {Phys. Rev. B}\ }\textbf {\bibinfo {volume} {67}},\ \bibinfo
  {pages} {235105} (\bibinfo {year} {2003})}\BibitemShut {NoStop}%
\bibitem [{\citenamefont {Fei}\ \emph {et~al.}(2018{\natexlab{b}})\citenamefont
  {Fei}, \citenamefont {Huang}, \citenamefont {Malinowski}, \citenamefont
  {Wang}, \citenamefont {Song}, \citenamefont {Sanchez}, \citenamefont {Yao},
  \citenamefont {Xiao}, \citenamefont {Zhu}, \citenamefont {May}, \citenamefont
  {Wu}, \citenamefont {Cobden}, \citenamefont {Chu},\ and\ \citenamefont
  {Xu}}]{Fei2018}%
  \BibitemOpen
  \bibfield  {author} {\bibinfo {author} {\bibfnamefont {Z.}~\bibnamefont
  {Fei}}, \bibinfo {author} {\bibfnamefont {B.}~\bibnamefont {Huang}}, \bibinfo
  {author} {\bibfnamefont {P.}~\bibnamefont {Malinowski}}, \bibinfo {author}
  {\bibfnamefont {W.}~\bibnamefont {Wang}}, \bibinfo {author} {\bibfnamefont
  {T.}~\bibnamefont {Song}}, \bibinfo {author} {\bibfnamefont {J.}~\bibnamefont
  {Sanchez}}, \bibinfo {author} {\bibfnamefont {W.}~\bibnamefont {Yao}},
  \bibinfo {author} {\bibfnamefont {D.}~\bibnamefont {Xiao}}, \bibinfo {author}
  {\bibfnamefont {X.}~\bibnamefont {Zhu}}, \bibinfo {author} {\bibfnamefont
  {A.~F.}\ \bibnamefont {May}}, \bibinfo {author} {\bibfnamefont
  {W.}~\bibnamefont {Wu}}, \bibinfo {author} {\bibfnamefont {D.~H.}\
  \bibnamefont {Cobden}}, \bibinfo {author} {\bibfnamefont {J.-H.}\
  \bibnamefont {Chu}}, \ and\ \bibinfo {author} {\bibfnamefont
  {X.}~\bibnamefont {Xu}},\ }\href {\doibase 10.1038/s41563-018-0149-7}
  {\bibfield  {journal} {\bibinfo  {journal} {Nature Materials}\ }\textbf
  {\bibinfo {volume} {17}},\ \bibinfo {pages} {778} (\bibinfo {year}
  {2018}{\natexlab{b}})}\BibitemShut {NoStop}%
\bibitem [{\citenamefont {Verchenko}\ \emph {et~al.}(2015)\citenamefont
  {Verchenko}, \citenamefont {Tsirlin}, \citenamefont {Sobolev}, \citenamefont
  {Presniakov},\ and\ \citenamefont {Shevelkov}}]{Verchenko2015}%
  \BibitemOpen
  \bibfield  {author} {\bibinfo {author} {\bibfnamefont {V.~Y.}\ \bibnamefont
  {Verchenko}}, \bibinfo {author} {\bibfnamefont {A.~A.}\ \bibnamefont
  {Tsirlin}}, \bibinfo {author} {\bibfnamefont {A.~V.}\ \bibnamefont
  {Sobolev}}, \bibinfo {author} {\bibfnamefont {I.~A.}\ \bibnamefont
  {Presniakov}}, \ and\ \bibinfo {author} {\bibfnamefont {A.~V.}\ \bibnamefont
  {Shevelkov}},\ }\href {\doibase 10.1021/acs.inorgchem.5b01260} {\bibfield
  {journal} {\bibinfo  {journal} {Inorganic Chemistry}\ }\textbf {\bibinfo
  {volume} {54}},\ \bibinfo {pages} {8598} (\bibinfo {year}
  {2015})}\BibitemShut {NoStop}%
\bibitem [{\citenamefont {Chen}\ \emph
  {et~al.}(2013{\natexlab{c}})\citenamefont {Chen}, \citenamefont {Yang},
  \citenamefont {Wang}, \citenamefont {Imai}, \citenamefont {Ohta},
  \citenamefont {Michioka}, \citenamefont {Yoshimura},\ and\ \citenamefont
  {Fang}}]{Bin2013}%
  \BibitemOpen
  \bibfield  {author} {\bibinfo {author} {\bibfnamefont {B.}~\bibnamefont
  {Chen}}, \bibinfo {author} {\bibfnamefont {J.~H.}\ \bibnamefont {Yang}},
  \bibinfo {author} {\bibfnamefont {H.~D.}\ \bibnamefont {Wang}}, \bibinfo
  {author} {\bibfnamefont {M.}~\bibnamefont {Imai}}, \bibinfo {author}
  {\bibfnamefont {H.}~\bibnamefont {Ohta}}, \bibinfo {author} {\bibfnamefont
  {C.}~\bibnamefont {Michioka}}, \bibinfo {author} {\bibfnamefont
  {K.}~\bibnamefont {Yoshimura}}, \ and\ \bibinfo {author} {\bibfnamefont
  {M.}~\bibnamefont {Fang}},\ }\href {\doibase 10.7566/JPSJ.82.124711}
  {\bibfield  {journal} {\bibinfo  {journal} {Journal of the Physical Society
  of Japan}\ }\textbf {\bibinfo {volume} {82}},\ \bibinfo {pages} {124711}
  (\bibinfo {year} {2013}{\natexlab{c}})}\BibitemShut {NoStop}%
\bibitem [{\citenamefont {Fisher}\ and\ \citenamefont
  {Langer}(1968)}]{Fisher1968}%
  \BibitemOpen
  \bibfield  {author} {\bibinfo {author} {\bibfnamefont {M.~E.}\ \bibnamefont
  {Fisher}}\ and\ \bibinfo {author} {\bibfnamefont {J.~S.}\ \bibnamefont
  {Langer}},\ }\href {\doibase 10.1103/PhysRevLett.20.665} {\bibfield
  {journal} {\bibinfo  {journal} {Phys. Rev. Lett.}\ }\textbf {\bibinfo
  {volume} {20}},\ \bibinfo {pages} {665} (\bibinfo {year} {1968})}\BibitemShut
  {NoStop}%
\bibitem [{\citenamefont {Mustafa}\ \emph {et~al.}(2016)\citenamefont
  {Mustafa}, \citenamefont {Bernardi}, \citenamefont {Neaton},\ and\
  \citenamefont {Louie}}]{Jamal2016}%
  \BibitemOpen
  \bibfield  {author} {\bibinfo {author} {\bibfnamefont {J.~I.}\ \bibnamefont
  {Mustafa}}, \bibinfo {author} {\bibfnamefont {M.}~\bibnamefont {Bernardi}},
  \bibinfo {author} {\bibfnamefont {J.~B.}\ \bibnamefont {Neaton}}, \ and\
  \bibinfo {author} {\bibfnamefont {S.~G.}\ \bibnamefont {Louie}},\ }\href
  {\doibase 10.1103/PhysRevB.94.155105} {\bibfield  {journal} {\bibinfo
  {journal} {Phys. Rev. B}\ }\textbf {\bibinfo {volume} {94}},\ \bibinfo
  {pages} {155105} (\bibinfo {year} {2016})}\BibitemShut {NoStop}%
\bibitem [{\citenamefont {Abrikosov}\ \emph {et~al.}(2016)\citenamefont
  {Abrikosov}, \citenamefont {Ponomareva}, \citenamefont {Steneteg},
  \citenamefont {Barannikova},\ and\ \citenamefont {Alling}}]{Abrikosov2016}%
  \BibitemOpen
  \bibfield  {author} {\bibinfo {author} {\bibfnamefont {I.}~\bibnamefont
  {Abrikosov}}, \bibinfo {author} {\bibfnamefont {A.}~\bibnamefont
  {Ponomareva}}, \bibinfo {author} {\bibfnamefont {P.}~\bibnamefont
  {Steneteg}}, \bibinfo {author} {\bibfnamefont {S.}~\bibnamefont
  {Barannikova}}, \ and\ \bibinfo {author} {\bibfnamefont {B.}~\bibnamefont
  {Alling}},\ }\href {\doibase https://doi.org/10.1016/j.cossms.2015.07.003}
  {\bibfield  {journal} {\bibinfo  {journal} {Current Opinion in Solid State
  and Materials Science}\ }\textbf {\bibinfo {volume} {20}},\ \bibinfo {pages}
  {85} (\bibinfo {year} {2016})}\BibitemShut {NoStop}%
\end{thebibliography}

%

\end{document}